\documentclass[journal]{IEEEtran}

\usepackage{float}
\usepackage{amssymb,amsmath,amsthm,bm}

\usepackage{empheq}
\AtBeginDocument{\renewcommand{\d}{\rm{d}}}

\usepackage{graphicx,color}
\graphicspath{{figures_pdf/}}
\usepackage{cite}
\usepackage{url}
\usepackage{diagbox}

\usepackage[aboveskip=1pt]{subcaption}
\usepackage{siunitx}
\usepackage{algorithm}
\usepackage{algpseudocode}
\usepackage{multirow,bigstrut}
\usepackage{tabularx}
\usepackage{threeparttable}
\usepackage{siunitx}

\usepackage{amsmath}
\DeclareMathOperator{\sign}{sign}

\newlength\OneImW
\newlength\TwoImW
\newlength\imagewidth
\newlength\figsep
\usepackage{hyperref}
\hypersetup{
 linktocpage=true, pdfborderstyle={/S/S/W 1}, hyperindex=true, bookmarks=true, bookmarksopen=true, bookmarksnumbered=true,
}

\usepackage{lineno}

 \usepackage{datetime}

\let\oldbibitem\bibitem
\def\bibitem{\vfill\oldbibitem}

\begin{document}

\title{Generating Multi-Scroll Chua's Attractors via\\
Simplified Piecewise-Linear Chua's Diode}

\author{Ning~Wang,~Chengqing~Li,~\IEEEmembership{Senior Member,~IEEE,}~Han~Bao,~Mo~Chen,~Bocheng~Bao
\thanks{This work was supported by the National Natural Science Foundation of China under Grant Nos. 51777016, 61601062, 61532020, and 61772447.}

\thanks{N. Wang, H. Bao, M. Chen, and B. Bao are with the School of Information Science and Engineering, Changzhou University, Changzhou 213164, China
(e-mail: cczuwangning@163.com, mervinbao@126.com).
}

\thanks{C. Li is with the College of Computer Science and Electronics Engineering, Hunan University, Changsha 410082, China (e-mail: DrChengqingLi@gmail.com).}}
\markboth{IEEE Transactions on Circuits and Systems I: Regular Papers}{Wang\MakeLowercase{et al.}}

\IEEEpubid{\begin{minipage}{\textwidth}\ \\[12pt] \centering
1549-8328 \copyright 2019 IEEE. Personal use is permitted, but republication/redistribution requires IEEE permission.\\
  See http://www.ieee.org/publications\_standards/publications/rights/index.html for more information.
\end{minipage}}

\maketitle
\begin{abstract}
High implementation complexity of multi-scroll circuit is a bottleneck problem in real chaos-based communication. Especially, in multi-scroll Chua's circuit, the simplified implementation of piecewise-linear resistors with multiple segments is difficult due to their intricate irregular breakpoints and slopes. To solve the challenge, this paper presents a systematic scheme for synthesizing a Chua's diode with multi-segment piecewise-linearity, which is achieved by cascading even-numbered passive nonlinear resistors with odd-numbered ones via a negative impedance converter. The traditional voltage mode op-amps are used to implement nonlinear resistors. As no extra DC bias voltage is employed, the scheme can be implemented by much simpler circuits. The voltage-current characteristics of the obtained Chua's diode are analyzed theoretically and verified by numerical simulations. Using the Chua's diode and a second-order active Sallen-Key high-pass filter, a new inductor-free Chua's circuit is then constructed to generate multi-scroll chaotic attractors. Different number of scrolls can be generated by changing the number of passive nonlinear resistor cells or adjusting two coupling parameters. Besides, the system can be scaled by using different power supplies, satisfying the low-voltage low-power requirement of integrated circuit design. The circuit simulations and hardware experiments both confirmed the feasibility of the designed system.
\end{abstract}
\begin{IEEEkeywords}
Chaos, Chua's diode, Chua's circuit, multi-scroll chaotic attractor, Secure communication.
\end{IEEEkeywords}

\section{Introduction}

\IEEEPARstart{S}{INCE} Leon O. Chua proposed the first third-order autonomous chaotic oscillator and investigated systems owning double scrolls in 1980's \cite{Chua1986The,Fortuna2009Chua},
the study of design, dynamics analysis and implementation of chaotic system has received much attention from the researchers in the field of nonlinear science, pseudo-random number generator, and chaotic cryptography \cite{Elwakil:ITCSITA:2000,Yalcin:ITCSIP:2004,Shen2014,Jia:ITCSIP:2014,Bao2017ASimple,Hua2018Sine,cqli:autoblock:IEEEM18}.
The subtle similarities between complex dynamics of chaotic systems and the basic requirements of a secure cryptosystem also promoted research of the topic \cite{cqli:IEAIE:IE18,Huazy:CAT:IEETC18,cqli:network:TCASI2019}. Using all kinds of enhancing or controlling methods, the systems with various topological properties were obtained in the past two decades: multi-scroll attractor \cite{JINHU2006GENERATING,maj:scro:PLOS18}, multi-wing chaotic system \cite{Elwakil2002Creation,A2003A,GUOYUAN2006FOUR}, hyperchaotic system \cite{Volos2017A}, multi-wing chaotic attractor \cite{Yu2008Generation, Huang2015Novel,Tahir2016A,Hong2018A} and hyperchaotic attractor \cite{Yu2012Design,Zhou2016Generating,Zhou2017A}.

Owing to the simple circuit structure and physical feasibility of Chua's circuit, it is considered as a prototype for investigating multi-scroll attractors and other nonlinear systems. In earlier 1990's, Suykens and Vandewalle designed the first $n$-double scroll Chua's attractors by using a so-called quasi-linear function \cite{Suykens1991Quasilinear,Suykens1993Generation}. In 1997, Suykens et al. proposed a more complete family of $n$-scroll Chua's attractors with the piecewise-linear function \cite{Suykens1997}. However, among these works, only numerical simulations were performed to verify existence of the multi-scroll attractors. In 2000, Yal\c{c}in et al. presented circuit implementation of 3- and 5-scroll attractors from a generalized Chua's circuit \cite{Yal2000Experimental}.
Adopting a sine or cosine function, Tang et al. modified Chua's circuit to generate  chaotic attractors owning $6\sim 9$ scrolls \cite{Tang2001Generation}. By using a multi-segment piecewise-linear Chua's diode, Zhong et al. proposed a systematic approach to generate multiple-scroll Chua's attractors owning up to 10 scrolls \cite{GUOQUNZHONG2002A}. Yu et al. presented an improved design approach for generating Chua's attractors with much more scrolls \cite{Yu2003New}. Especially, the sizes of the obtained scrolls with the method proposed in \cite{Yu2003New} are uniform. In 2007, Yu et al. further extended their methods to
generate $n\times m$-scroll attractors from a \textit{single} Chua's circuit using a sawtooth function and a staircase function \cite{SIMIN2007GENERATION}.
Multi-direction multi-double-scroll Chua's attractors can be obtained by constructing pulsed excitation in the three dimensions of the corresponding state variable \cite{Hong2017A}. In \cite{Wang2017Multi}, Wang
et al. generated multi-scroll attractors from the memristive Chua's circuit by introducing memristors with a multi-piecewise continuous memductance function. In \cite{Karthikeyan2018NOVEL}, a parametrically controlled method was used to generate multi-scroll Chua's attractors.

\IEEEpubidadjcol 

The implementation complexities of the above reviewed construction methods of multi-scroll attractors vary in a large range. Some of them involve a large number of discrete components \cite{GUOQUNZHONG2002A,Yu2003New,Wang2017Multi,Karthikeyan2018NOVEL}, extra DC bias voltages \cite{GUOQUNZHONG2002A,Tahir2016A,Zhou2016Generating,Wang2017Multi,Karthikeyan2018NOVEL}, or pulse excitations \cite{Hong2017A,Hong2018A,Wang2018Emerging}.
Such elements inevitably lead to complex selection of parasitic parameters, waste of hardware resources and cost of a large amount of energy.
However, low-power consumption and facilitation are the basic requirements in some application scenarios, e.g. IoT devices and wireless communication \cite{Jin2018Low}.
What's more, the electronic circuit with manually winding inductor element is bulky in terms of size and unsuitable for IC design.
To address the challenges, some inductor-free Chua's circuits were designed, e.g. inductance simulator based on general impedance converter \cite{Torres2000Inductorless}, Chua's circuit based on electronic analogy \cite{Rocha2009An},
Chua's circuit based on Wien bridge oscillator \cite{Morgul2002Inductorless} and Chua's circuit variants
based on active band pass filter \cite{Banerjee2012Single,Bao2016Inductor}.

The design and implementation of simple autonomous chaotic circuits with complex topological structure are important for theoretical analysis and practical applications of chaotic signals. This paper focuses on the following two points: 1) simplifying the implementation of multi-piecewise Chua's diode and providing a systematic way to determine element parameters; 2) extending the multi-scroll and inductor-free Chua's circuit family. As Chua's diode is the part of Chua's circuit consuming most energy of the whole system, changing its generation mechanism is an effective way to obtain multi-scroll attractors with low circuit complexity. The saturation characteristic of op-amp is critical for the formulation of breakpoints. Obviously, the cascading combination of multiple nonlinear resistor cells is an approach for constructing a multi-segment piecewise-linear Chua's diode. Based on this idea, we developed a simple scheme implementing a multi-segment piecewise-linear Chua's diode and an inductor-free multi-scroll Chua's circuit. Both theoretical analyses and experimental simulations were provided to validate desired performances of the modified Chua's system.

The remainder of the paper is organized as follows. In Sec.~\ref{sec:SimpliChuadiode}, we introduce a scheme for implementing a multi-segment piecewise-linear Chua's diode. A third-order inductor-free Chua's circuit
based on active Sallen-Key high-pass filter (HPF) is designed in Sec.~\ref{sec:Chuacircuit}. An encryption application based on the proposed multi-scroll attractor is presented in Sec.~\ref{sec:Application}.
The multisim circuit simulations
and the hardware experiment results of the circuit are provided in Sec.~\ref{sec:Experiments}. The last section concludes the paper.

\section{Simplified Piecewise-linear Chua's diode}
\label{sec:SimpliChuadiode}

In this section, we first perform circuit syntheses of two types of nonlinear resistors, and then give a design scheme for implementing a simplified multi-segment piecewise-linear Chua's diode. Finally, two $v-i$ characteristics with 17 segments and 19 segments are presented.

\subsection{Two Types of Nonlinear Resistors}
\label{subsec:NonlinearResistors}

The scheme adopted here is to combine multiple op-amp-based nonlinear resistors owning different breakpoints. Assume the input currents $i_{\rm +}$ and $i_{\rm -}$ of the voltage mode op-amps (VOAs) are equal to zero. The potential difference between two input terminals is $v_D$. The gain of the op-amp in the linear region tends to infinity. Two types of nonlinear resistors, named as passive 3-segment piecewise-linear resistor and active one, can be implemented by two sets of op-amp-based circuits, respectively.
To facilitate description, the passive 3-segment piecewise-linear resistor is defined as type-I nonlinear resistor marked by $N_{\rm RI}$. The active 3-segment piecewise-linear resistor, by contrast, is defined as type-II nonlinear resistor marked by $N_{\rm RII}$.
Then, the two types of nonlinear resistors can be analyzed as follows.

\subsubsection{Type-I Passive Nonlinear Resistor}

\begin{figure}[!htb]
	\centering
	\begin{minipage}{\OneImW}
		\centering
		\includegraphics[width=0.93\OneImW]{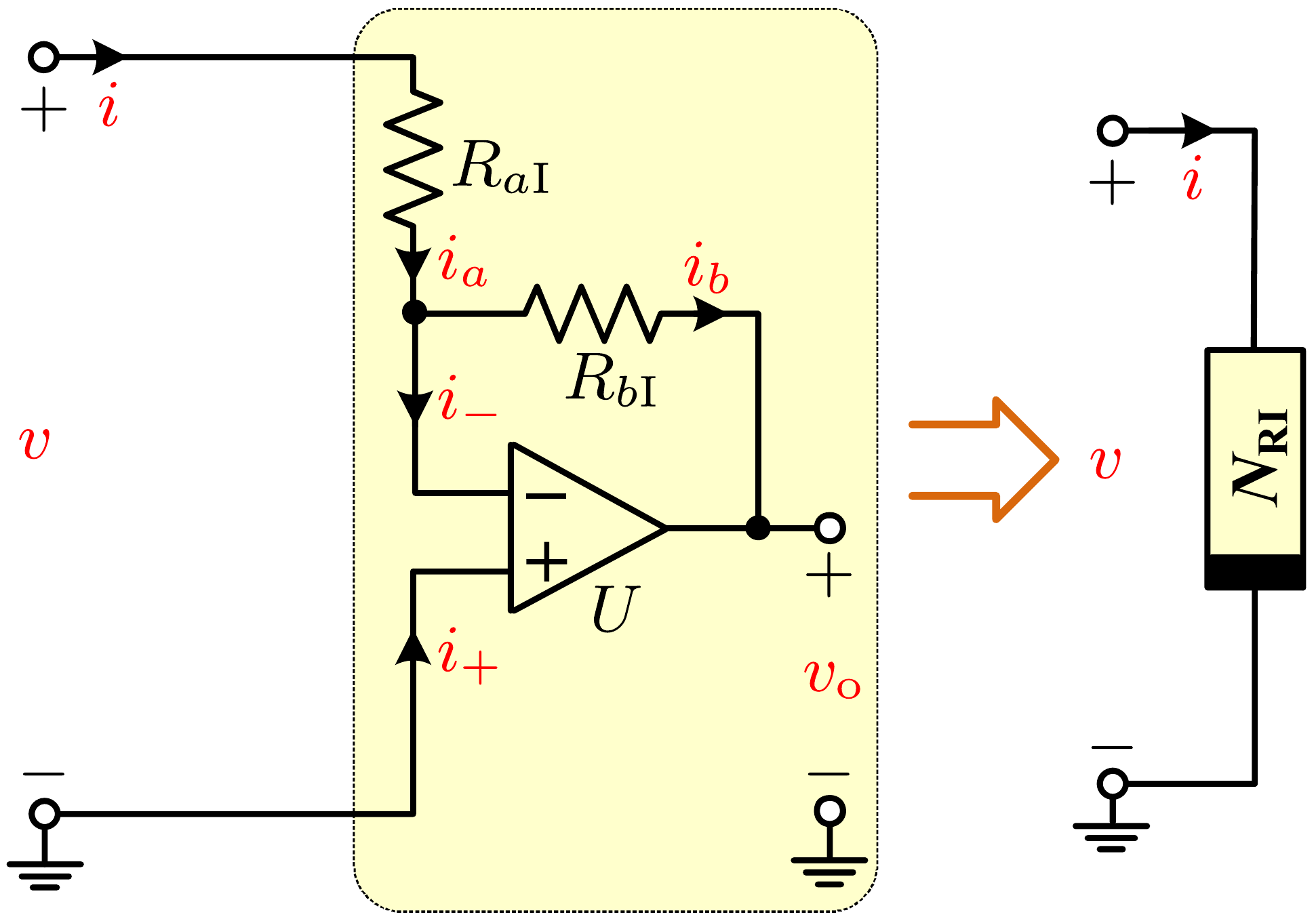}
		(a)
	\end{minipage}\\
	\begin{minipage}{\OneImW}
		\centering
		\includegraphics[width=\OneImW]{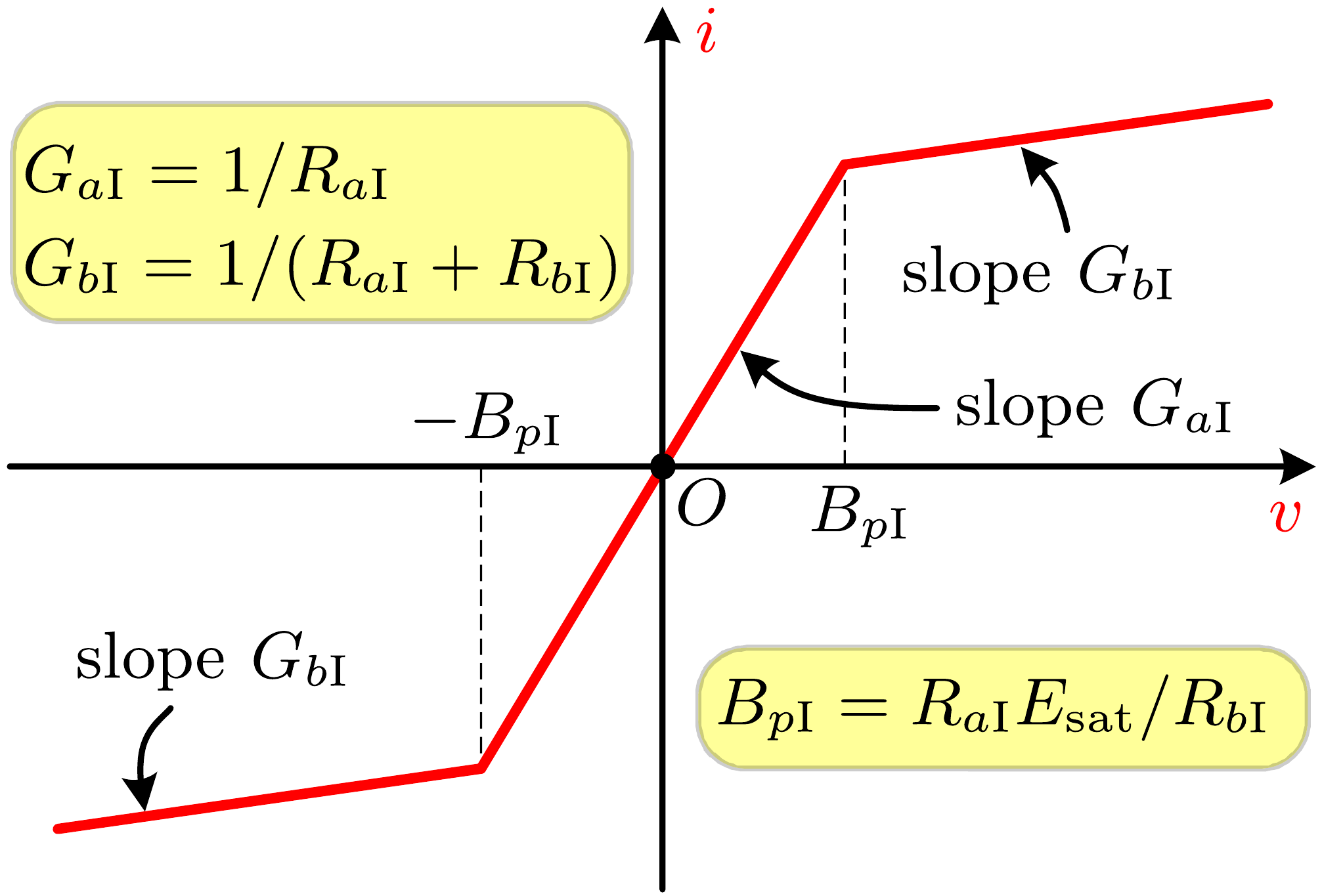}
		(b)
	\end{minipage}
	\caption{Design of the type-I passive nonlinear resistor $N_{\rm RI}$: (a) circuit implementation; (b) the $v-i$ characteristic curve.}
\label{fig:NRI}
\end{figure}

The schematic diagram of the type-I nonlinear resistor is illustrated in Fig.~\ref{fig:NRI}(a), which contains one op-amp and two resistors. Obviously, it has the same circuit topology as the op-amp based gain circuit.

By applying Kirchhoff's voltage and current laws, the voltage-current characteristic of $N_{\rm RI}$ can be classified as the following three cases:

\begin{itemize}
\item $|v|\leq B_{p\rm I}$:

In this case, the op-amp works in a linear region. So, $\pm B_{p\rm I}=\pm R_{a\rm I}E_{\rm sat}/R_{b\rm I}$ are two breakpoints at both ends of the linear region. Then, one has $v_D= 0 $, $i=(v - v_D)/R_{a\rm I}$. Thus, the inner slope
is
\begin{equation*}
G_{a\rm{I}}=\frac{i}{v}=\frac{1}{R_{a{\rm I}}}.
\end{equation*}

\item $v>B_{p\rm I}$:

Under such condition, the op-amp works in the negative saturation region and $v_{\rm o}=-E_{\rm sat}$.
According to the current flowing through $N_{\rm RI}$  (i.e., the current passing $R_{a\rm I}$ and $R_{b\rm I}$), one can get
\begin{equation}
\frac{v-{v_D}}{R_{a{\rm I}}}=\frac{v_D-v_{\rm o}}{R_{b{\rm I}}},
\label{eq:iaib}
\end{equation}
where ${v_D}\ne 0$.
Then, one can obtain
\begin{equation}
v_D=\frac{R_{b{\rm I}}v-R_{a{\rm I}}E_{\rm sat}}{R_{a{\rm I}}+R_{b{\rm I}}}.
\label{eq:vD}
\end{equation}
Substituting~(\ref{eq:vD}) into~(\ref{eq:iaib}), the current function related to the input voltage $v$ can be presented as
\begin{equation*}
 i=f(v)=\frac{v+E_{\rm sat}}{R_{a{\rm I}}+R_{b{\rm I}}}.
\end{equation*}

At the breakpoint, $B_{p\rm I}=R_{a\rm I}E_{\rm sat}/R_{b\rm I}$ and $G_{a\rm I}B_{p\rm I}= E_{\rm sat}/R_{b\rm I}$. Thus, the outer slope can be calculated as
\begin{equation*}
 G_{b{\rm I}}=\frac{\Delta i}{\Delta v}=\frac{f(v)-G_{a{\rm I}}B_{p{\rm I}}}{v-B_{p{\rm I}}}=\frac{1}{R_{a{\rm I}}+R_{b{\rm I}}}.
\end{equation*}

\item $v<-B_{p\rm I}$:
In this case, the op-amp works in positive saturation region and $v_{\rm o}=E_{\rm sat}$. The outer slope can be derived as $G_{b\rm I}=1/(R_{a\rm I}+R_{b\rm I})$.
\end{itemize}

According to the above three cases, the $v-i$ characteristic of the $N_{\rm RI}$ is drawn as shown in Fig.~\ref{fig:NRI}(b), which can be featured by three parameters and represented as a uniform form
\begin{align*}
i & =h(v)  \\
& =G_{b{\rm I}}v+0.5(G_{a{\rm I}}-G_{b{\rm I}})(|v+B_{p{\rm I}}|-|v-B_{p{\rm I}}|),
\end{align*}
where
\begin{equation*}
G_{a{\rm I}}=\frac{1}{R_{a{\rm I}}},G_{b{\rm I}}=\frac{1}{R_{a{\rm I}}+R_{b{\rm I}}},B_{p{\rm I}}=\frac{R_{a{\rm I}}E_{\rm{sat}}}{R_{b{\rm I}}}
\end{equation*}
denote the inner slope, outer slope, and breakpoint, respectively. As the three parameters satisfy $G_{a\rm I}>0$, $G_{b\rm I}>0$, and $G_{a\rm I}>G_{b\rm I}$, the $v-i$ curve always lies in the first and third quadrants, and the inflow of power is always nonnegative. According to the definition of passivity \cite{khalil2002nonlinear}, the nonlinear resistor cell is passive.
The implementation occupies only one op-amp and two resistors. So the proposed type-I nonlinear resistor is the simplest op-amp-based piecewise-linear resistor.

\subsubsection{Type-II Active Nonlinear Resistor}

\begin{figure}[!htb]
	\centering
	\begin{minipage}{\OneImW}
		\centering
		\includegraphics[width=\OneImW]{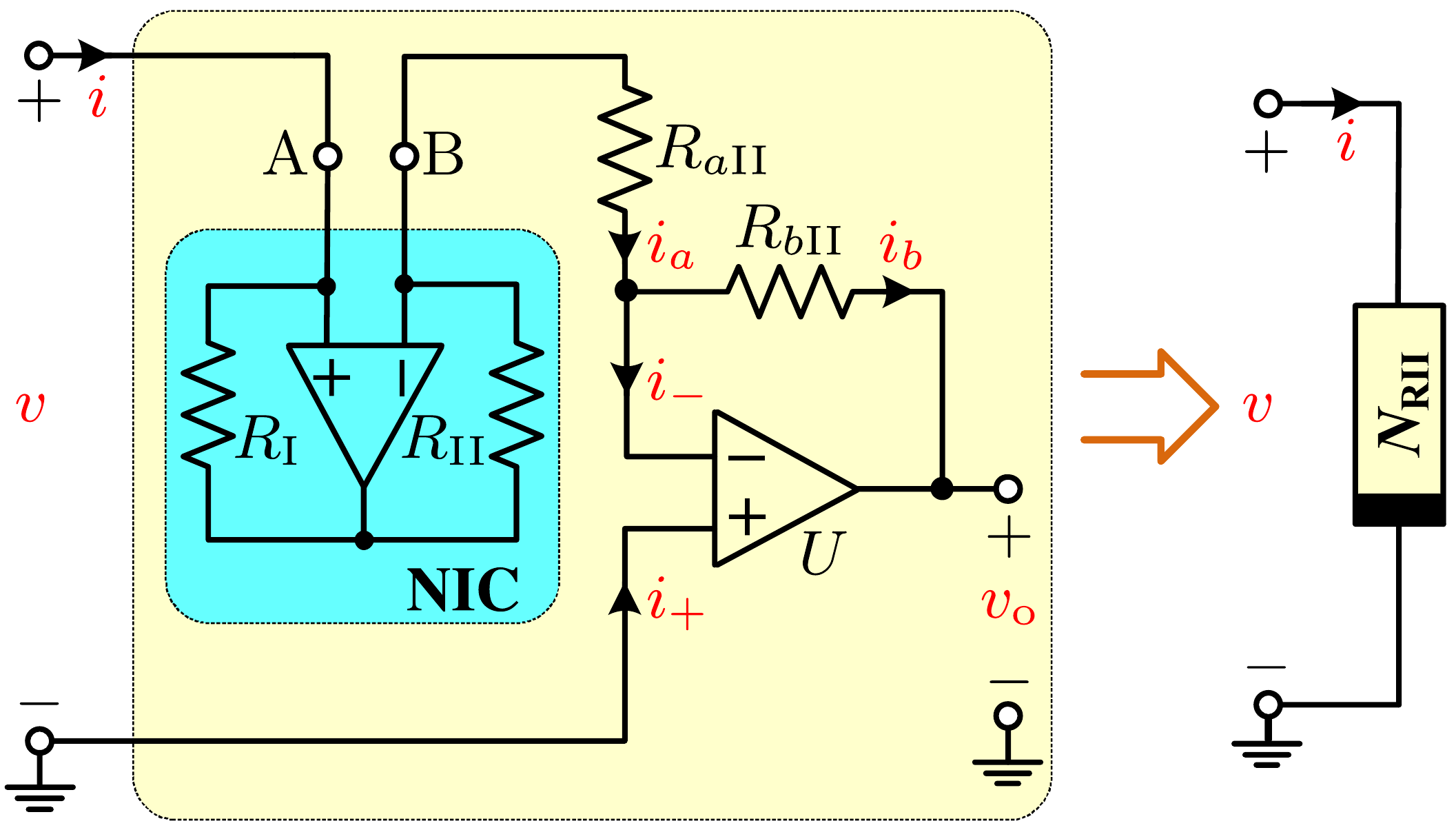}
		(a)
	\end{minipage}\\
	\begin{minipage}{\OneImW}
		\centering
		\includegraphics[width=\OneImW]{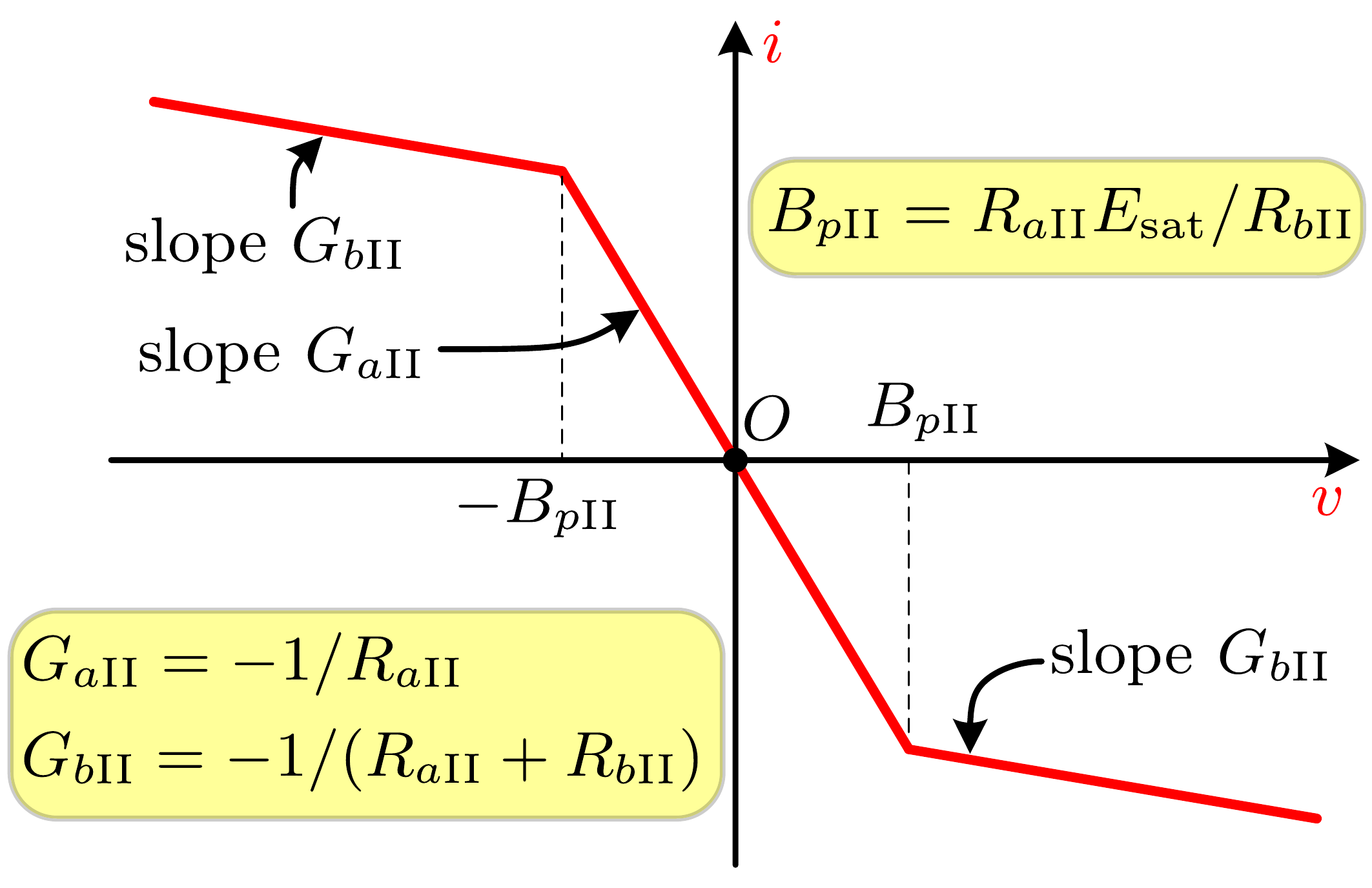}
		(b)
	\end{minipage}
	\caption{Design of the type-II active nonlinear resistor $N_{\rm RII}$: (a) circuit implementation; (b) the $v-i$ characteristic curve.}
	\label{fig:NRII}
\end{figure}

Linking a negative impedance converter (NIC) to the input port of the $N_{\rm RI}$, another type of active 3-segment piecewise-linear resistor can be obtained as shown in Fig.~\ref{fig:NRII}(a). The corresponding voltage-current characteristic of $N_{\rm RII}$ is shown in Fig. \ref{fig:NRII}(b), which can be described as
\begin{align*}
i & = h(v)\\
  &= G_{b{\rm II}}v+0.5(G_{a{\rm II}}-G_{b{\rm II}})(|v+B_{p{\rm II}}|-|v-B_{p{\rm II}}|),
\end{align*}
where $G_{a\rm II}$, $G_{b\rm II}$, and $B_{p\rm II}$ are the inner slope, outer slope, and breakpoint, respectively. When $R_{\rm I}=R_{\rm II}$, one has
\begin{equation*}
G_{a{\rm II}}=-\frac{1}{R_{a{\rm II}}},G_{b{\rm II}}=-\frac{1}{R_{a{\rm II}} + R_{b{\rm II}}},B_{p{\rm II}}=\frac{R_{a{\rm II}}E_{\rm{sat}}}{R_{b{\rm II}}}.
\end{equation*}

Observing Fig.~\ref{fig:NRII}, one can see that $G_{a\rm II}<0$, $G_{b\rm II}<0$, and $|G_{a\rm II}|>|G_{b\rm II}|$. Therefore, $N_{\rm RII}$ has the same characteristic as the classical Chua's diode with negative inner slope and outer slope. But, it employs fewer (four) resistors than the six ones in the classical Chua's diode designed in \cite{Fortuna2009Chua}.

\begin{figure}[!htb]
	\centering
	\includegraphics[width=\OneImW]{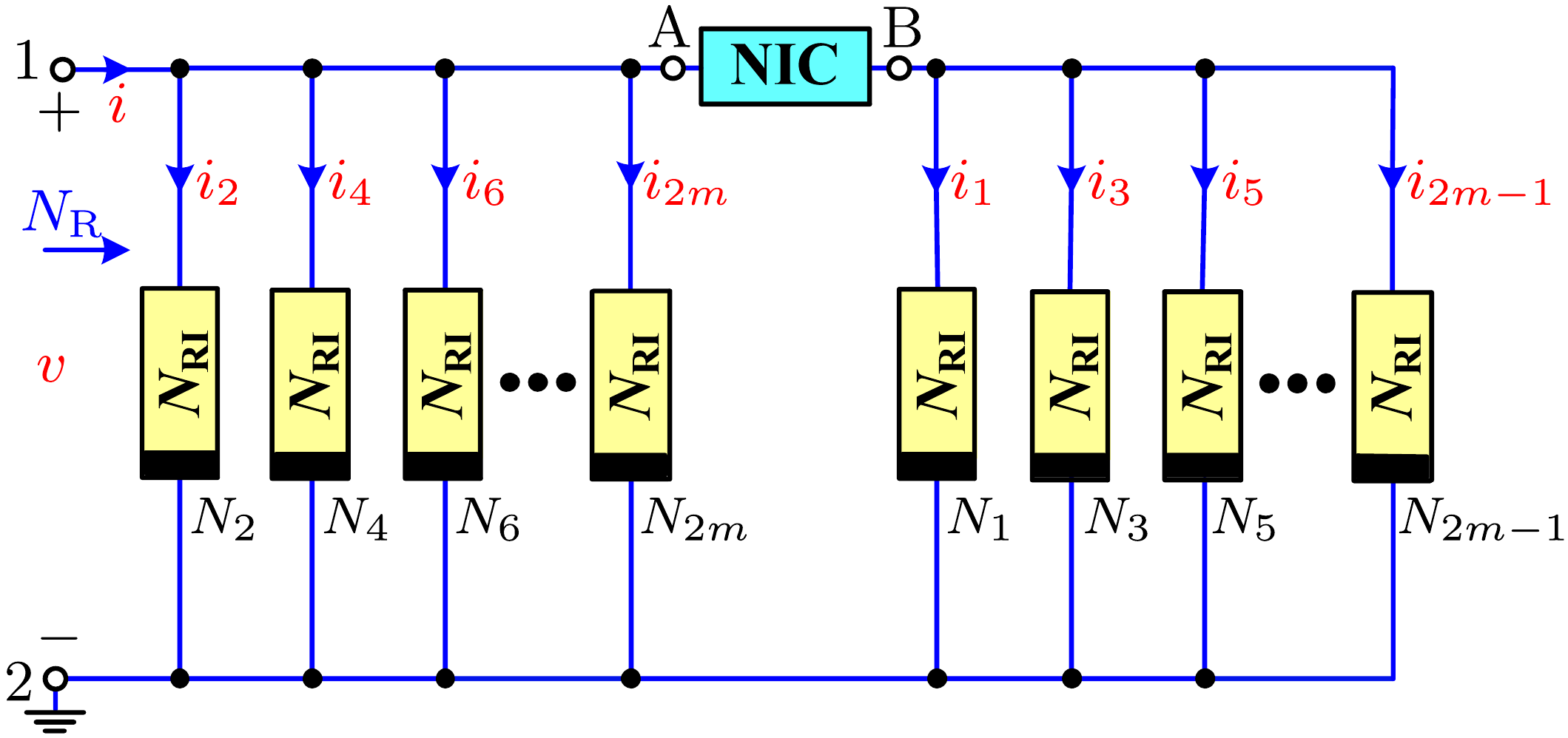}
	\caption{The schematic of multi-piecewise Chua's diodes, where the $m$-th $N_{\rm RI}$ is marked as $N_m$.}
	\label{fig:Chuadiode}
\end{figure}

\subsection{Multi-Segment Piecewise-Linear Chua's Diode}
\label{subsec:ChuaDiode}

To design the $v-i$ curves with multiple segments, some additional breakpoints need to be introduced \cite{Fortuna2009Chua}. Adjusting the linear resistances of the equivalent circuit of $N_{\rm RI}$, the inner slope, the outer slope and breakpoints can be freely adjusted. As shown in Fig.~\ref{fig:Chuadiode}, a scheme synthesizing the multi-segment piecewise-linear Chua's diode is proposed, where $N_{\rm RI}$ have the same inner slope but different breakpoints. Note that the NIC works as a current reverser here, which means that the odd-numbered $N_{\rm RI}$ belongs to type-II nonlinear resistors. Thus, the design scheme can also be regarded as the cascading combination of multiple $N_{\rm RI}$ and $N_{\rm RII}$ in turn.

As for the input voltage $v$ and the output current $i$ of the proposed Chua's diode, the voltage-current characteristic can be described as
\begin{multline}
h(v)=\sum\limits_{m=1}^M(-1)^m[G_{bm}v+0.5(G_{am}-G_{bm})\\
\times (|v+B_{pm}|-|v-B_{pm}|)],
\label{eq:vi}
\end{multline}
where
\begin{equation*}
G_{am}=\frac{1}{R_{am}},G_{bm}=\frac{1}{R_{am}+R_{bm}}, \mbox{and } B_{pm}=\frac{R_{am}E_{\rm{sat}}}{R_{bm}}
\end{equation*}
denote the inner slope, outer slope, and breakpoint of $N_m$, respectively.

Note that the $m$-th $N_{\rm RI}$, $N_m$, can generate two symmetric breakpoints $\pm B_{pm}$, which are determined by resistances $R_{am}$, $R_{bm}$ and op-amp saturation voltage $E_{\rm sat}$. Thus the cascading combination of $N_{\rm RI}$ with $2M$ breakpoints can generate $2M+1$ segments. The parameters
of $M$-$N_{\rm RI}$-based Chua's diode can be determined with the following steps:
\begin{itemize}
\item Keep the op-amp saturation voltage $E_{\rm sat}$ unchanged, and set the values of the outermost breakpoints to $\pm B_{p1}$.

\item Divide the interval between $-B_{p1}$ and $B_{p1}$ into $2M-1$ segments of equal intervals. The breakpoints $\pm B_{pm}$, generated from $N_m$, are calculated as
\begin{equation}
\pm B_{pm}=\pm \frac{2B_{p1}}{2M-1}(M-m+0.5),
\label{eq:Bpm}
\end{equation}
where $m=1, 2, \cdots, M$.

\item Fix the inner slope of $ N_{\rm RI}$, i.e., keep resistance $R_{a1}=R_{a2}=\cdots =R_{aM}$ unchanged.

\item Set \[R_{bm}=\frac{R_{am}}{B_{pm}}E_{\rm sat}.\]
\end{itemize}

\addtolength{\abovecaptionskip}{0pt}
\renewcommand{\arraystretch}{1.05}

\begin{table}[!htb]
	\caption{ Element Values and Parameters of Chua's Diode.}
	\centering
	\begin{tabular}{c|l|l}
		\hline
		$M$ & Element values & Parameters of Chua's diode\\
		\hline
		\multirow{8}{*}{8}
		&$R_{a\rm i}=\SI{1}{\kohm}$, $\rm i=1\sim 8$ & $G_{a\rm i} =\SI{1}{\milli\siemens}$,  $\rm i=1\sim 8$\\
		&$R_{b\rm 1}=\SI{3.25}{\kohm}$    &  $G_{b\rm 1}=\SI{235.294}{\micro\siemens}$, $B_{p\rm 1}=\SI{4}{\volt}$\\
		&$R_{b\rm 2}=\SI{3.75}{\kohm}$    &  $G_{b\rm 2}=\SI{210.526}{\micro\siemens}$,$B_{p\rm 2}=\SI{52/15}{\volt}$\\
		&$R_{b\rm 3}=\SI{4.432}{\kohm}$   &  $G_{b\rm 3}=\SI{184.094}{\micro\siemens}$,$B_{p\rm 3}=\SI{44/15}{\volt}$\\
		&$R_{b\rm 4}=\SI{5.417}{\kohm}$   &  $G_{b\rm 4}=\SI{155.836}{\micro\siemens}$,$B_{p\rm 4}=\SI{12/5}{\volt}$\\
		&$R_{b\rm 5}=\SI{6.964}{\kohm}$   &  $G_{b\rm 5}=\SI{125.565}{\micro\siemens}$,$B_{p\rm 5}=\SI{28/15}{\volt}$\\
		&$R_{b\rm 6}=\SI{9.75}{\kohm}$    &  $G_{b\rm 6}=\SI{93.023}{\micro\siemens}$,$B_{p\rm 6}=\SI{4/3}{\volt}$\\
		&$R_{b\rm 7}=\SI{16.25}{\kohm}$   &  $G_{b\rm 7}=\SI{57.971}{\micro\siemens}$,$B_{p\rm 7}=\SI{4/5}{\volt}$\\
		&$R_{b\rm 8}=\SI{48.75}{\kohm}$   &  $G_{b\rm 8}=\SI{20.101}{\micro\siemens}$,$B_{p\rm 8}=\SI{4/15}{\volt}$\\ \hline
		\multirow{9}{*}{9}
		& $R_{a\rm j} =\SI{1}{\kohm}$,  $\rm j=1\sim 9$ & $G_{a\rm j}=\SI{1}{\milli\siemens}$, $\rm j=1\sim 9$ \\
		& $R_{b\rm 1}=\SI{3.25}{\kohm}$                 & $G_{b\rm 1}=\SI{235.294}{\micro\siemens}$, $B_{p \rm 1}=\SI{4}{\volt}$\\
		& $R_{b\rm 2}=\SI{3.683}{\kohm}$                & $G_{b\rm 2}=\SI{213.538}{\micro\siemens}$, $B_{p \rm 2}=\SI{60/17}{\volt}$\\
		& $R_{b\rm 3}=\SI{4.25}{\kohm}$                 & $G_{b\rm 3}=\SI{190.476}{\micro\siemens}$, $B_{p \rm 3}=\SI{52/17}{\volt}$\\
		& $R_{b\rm 4}=\SI{5.023}{\kohm}$                & $G_{b\rm 4}=\SI{166.030}{\micro\siemens}$, $B_{p \rm 4}=\SI{44/17}{\volt}$\\
		& $R_{b\rm 5}=\SI{6.139}{\kohm}$                & $G_{b\rm 5}=\SI{140.076}{\micro\siemens}$, $B_{p \rm 5}=\SI{36/17}{\volt}$\\
		& $R_{b\rm 6}=\SI{7.893}{\kohm}$                & $G_{b\rm 6}=\SI{112.448}{\micro\siemens}$, $B_{p \rm 6}=\SI{28/17}{\volt}$\\
		& $R_{b\rm 7}=\SI{11.05}{\kohm}$                & $G_{b\rm 7}=\SI{82.988}{\micro\siemens}$, $B_{p \rm 7}=\SI{20/17}{\volt}$\\
		& $R_{b\rm 8}=\SI{18.417}{\kohm}$               & $G_{b\rm 8}=\SI{51.501}{\micro\siemens}$, $B_{p \rm 8}=\SI{12/17}{\volt}$\\
		& $R_{b\rm 9}=\SI{55.25}{\kohm}$                & $G_{b\rm 9}=\SI{17.778}{\micro\siemens}$, $B_{p \rm 9}=\SI{4/17}{\volt}$\\ \hline
	\end{tabular}
	\label{table:parameters}
\end{table}

\begin{figure}[!htb]
	\centering
	\begin{minipage}{\OneImW}
		\centering
		\includegraphics[width=\OneImW]{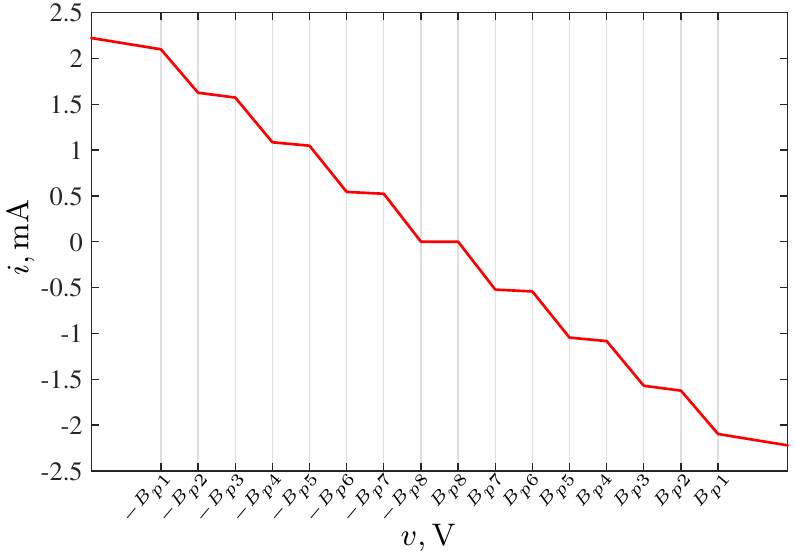}
		(a)
	\end{minipage}\\
	\begin{minipage}{\OneImW}
		\centering
		\includegraphics[width=\OneImW]{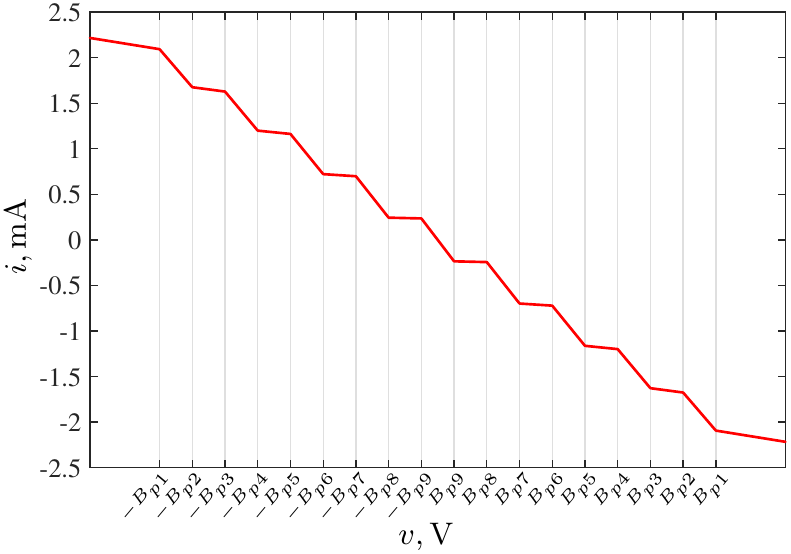}
		(b)
	\end{minipage}
	\caption{The $v-i$ characteristics of the proposed Chua's diodes: (a) the 17-segment piecewise-linear curve; (b) the 19-segment piecewise-linear curve.}
	\label{fig:simulatedvicurve}
\end{figure}

\subsection{The $v$--$i$ Characteristic of the Proposed Chua's Diode}
\label{subsec:ChuaDiodevi}

To illustrate the $v$-$i$ characteristics of the proposed Chua's diodes, the 8-$N_{\rm RI}$ and 9-$N_{\rm RI}$-based Chua's diodes are used as two examples. The resistances $R_{am}$ are fixed as \SI{1}{\kohm} for $m=1\sim M$. In addition, the op-amp saturation output voltage $E_{\rm sat}=\SI{ 13}{\volt}$ and the outermost breakpoints $\pm B_{p1}=\SI{\pm 4}{\volt}$ are adopted. Using the parameter determination steps described in the above sub-section, two sets of element values are calculated and listed in Table~\ref{table:parameters}. The loci of the two Chua's diodes in the $v-i$ plane are plotted in Fig.~\ref{fig:simulatedvicurve}(a) and (b), which depict 17 segments and 19 segments, respectively. Considering that a NIC-based current reverser is necessary, the two examples can be also named as 9-stage and 10-stage op-amp-based Chua's diodes, respectively.

Comparing with the construction approach of $n$-scroll attractor in \cite{GUOQUNZHONG2002A}, all extra DC bias voltages are removed in the proposed approach. The reduction or elimination of external DC voltage references is of vital importance for simplifying circuit implementations, especially for the integrated circuits \cite{Trejo2012Integrated}. Meanwhile, fewer resistors are employed for generating the same number of segments. Taking 17 segments as an example, the proposed Chua's diode employs 18 resistors.
In contrast, the numbers required in the construction methods proposed in \cite{GUOQUNZHONG2002A} and \cite{Yu2003New} are 27 and 48, respectively.

\begin{figure}[!htb]
	\centering
	\includegraphics[width=\OneImW]{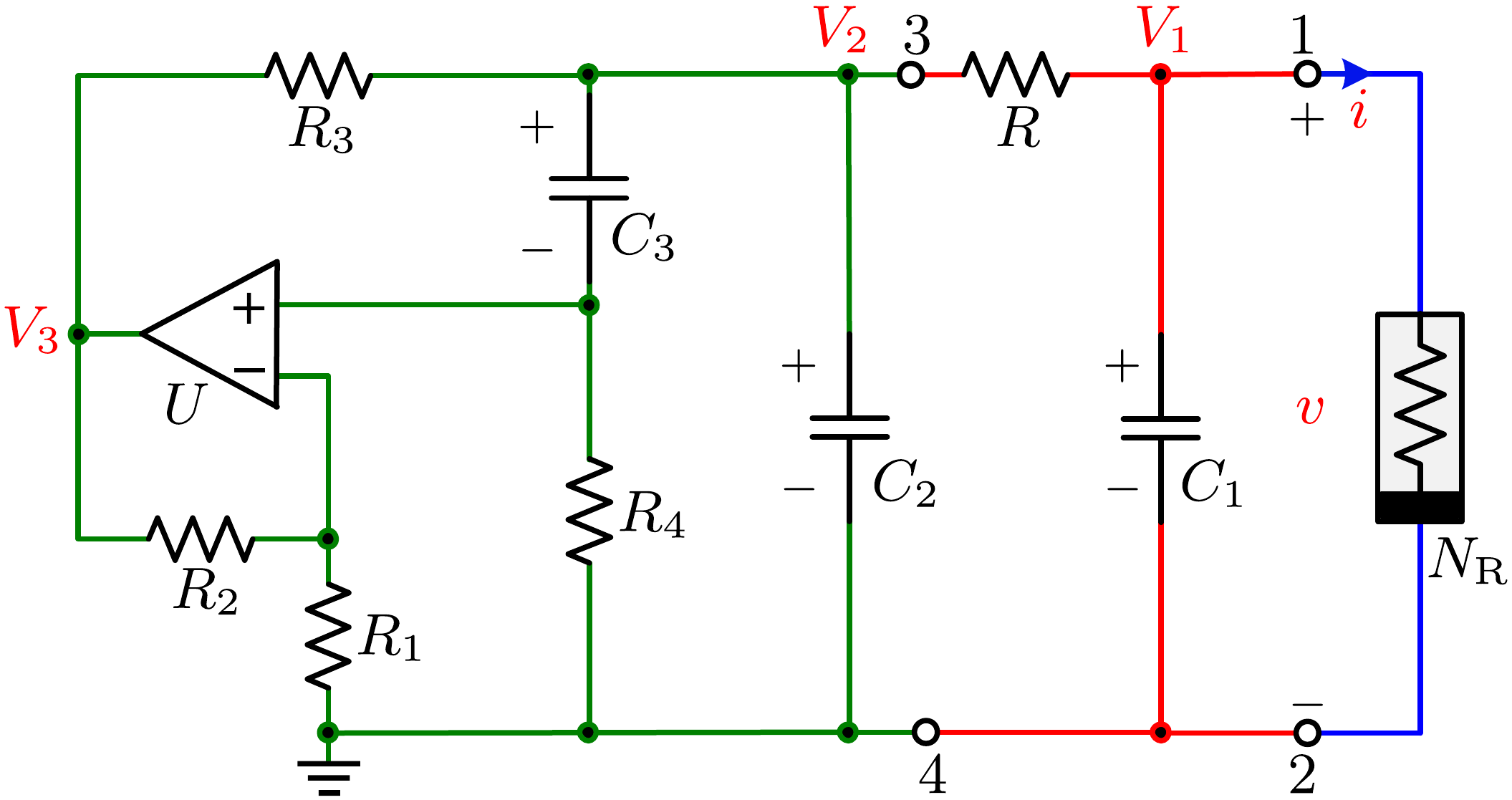}
	\caption{Schematic of the Sallen-Key HPF-based Chua's circuit.}
	\label{fig:Circuitschematic}
\end{figure}

\section{Sallen-Key HPF-Based Chua's Circuit}
\label{sec:Chuacircuit}

In this section, a third-order inductor-free Chua's circuit is designed to generate multi-scroll chaotic attractors
using the proposed Chua's diode.

\subsection{Ciruit Description and State Equation}
\label{subsec:CiruitDescription}

As shown in Fig.~\ref{fig:Circuitschematic}, an inductor-free Chua's circuit is constructed by replacing the parallel connection LC network in the classical Chua's circuit with a second-order active Sallen-Key high-pass filter (HPF) designed in \cite{Van1982Analog}. The proposed Chua's diode is placed in the right part of the ports ``1'' and ``2''. Meanwhile, a second-order Sallen-Key HPF is set in the left part of the ports ``3'' and ``4''. In the Sallen-Key HPF, $C_{\rm 2}=C_{\rm 3}$ and $R_{\rm 3}=R_{\rm 4}$. Comparing with the Wien bridge oscillator\cite{Morgul2002Inductorless}, second-order active band pass filter \cite{Banerjee2012Single,Bao2016Inductor}, and second-order active Sallen-Key low-pass filter \cite{Bao2017Chaotic}, the Sallen-Key HPF has the same off-the-shelf discrete components but different circuit structures.

The circuit shown in Fig.~\ref{fig:Circuitschematic} is a third-order autonomous circuit containing three dynamical elements of capacitors $C_{\rm 1}$, $C_{\rm 2}$, and $C_{\rm 3}$, which can be featured by three coupled first-order autonomous differential equations in terms of node voltages $V_{\rm 1}$, $V_{\rm 2}$, and $V_{\rm 3}$,
\begin{empheq}[left=\empheqlbrace]{align}
\frac{{\d}V_1}{{\d}t}&=-\frac{V_1-V_2}{RC_1}-\frac{h(V_1)}{C_1}, \nonumber\\
\frac{{\d}V_2}{{\d}t}&=\frac{V_1-V_2}{RC_2}+\frac{(k-1)V_3-kV_2}{k R_3C_2} \label{eq:circuitequation2},\\
\frac{{\d}V_3}{{\d}t}&=\frac{k(V_1-V_2)}{RC_3}+\frac{(k-2)V_3-kV_2}{R_3C_3}, \nonumber
\end{empheq}
where $k=1+\frac{R_2}{R_1}$, and $h(V_{\rm 1})$ generated by Eq.~(\ref{eq:vi}) stands for the voltage-current characteristic of the Chua's diode.

\begin{figure}[!htb]
	\centering
	\begin{minipage}{\OneImW}
		\centering
		\includegraphics[width=\OneImW]{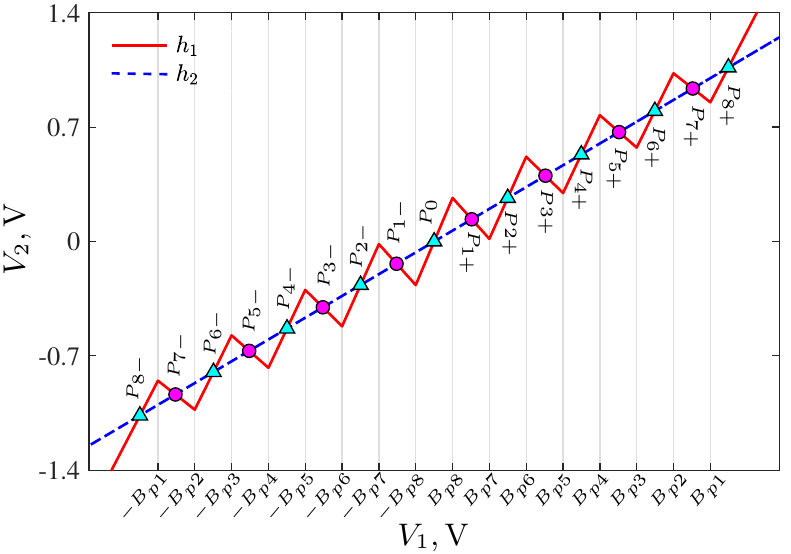}
		(a)
	\end{minipage}\\
	\begin{minipage}{\OneImW}
		\centering
		\includegraphics[width=\OneImW]{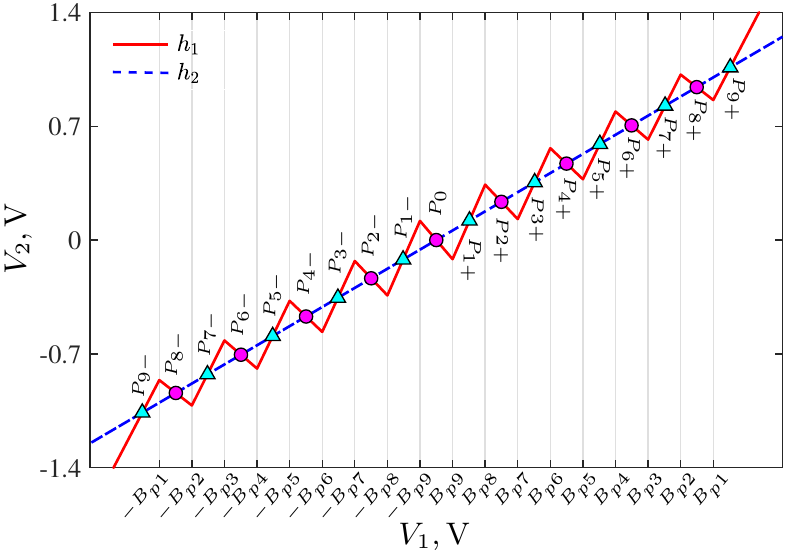}
		(b)
	\end{minipage}
	\caption{Equilibrium points in the $V_1-V_2$ plane: (a) Chua's circuit owing 17 equilibrium points. (b) Chua's circuit owing 19 equilibrium points.}
	\label{fig:Equilibriumpoints}
\end{figure}

\renewcommand{\arraystretch}{1.05}
\begin{table*}[!htb]
	\caption{Equilibrium Points and the Corresponding Eigenvalues.}
	\centering
	\begin{tabular}{c|l|l|l}
		\hline
		$M$ & Equilibrium points &  Eigenvalues & Types of equilibrium points\\
		\hline
		\multirow{8}{*}{8} & $P_0:(\SI{0}{\volt}, \SI{0}{\volt}, \SI{0}{\volt})$ & $\lambda_{1,2}=1162\pm j19475,\lambda_3=-103030$ & unstable index-2 saddle-focus\\ \cline{2-4}
		& $P_{1\pm}:(\SI{\pm 0.5445}{\volt},\SI{\pm 0.1361}{\volt},\SI{0}{\volt})$ & $\lambda_1=58532,\lambda_{2,3}=-7569\pm j24202$ & unstable index-1 saddle-foci\\ \cline{2-4}
		& $P_{2\pm}:(\SI{\pm 1.0653}{\volt},\SI{\pm 0.2663}{\volt},\SI{0}{\volt})$ & $\lambda_{1,2}=1254\pm j19226,\lambda_3=-97644$ & unstable index-2 saddle-foci\\ \cline{2-4}
		& $P_{3\pm}:(\SI{\pm 1.6118}{\volt},\SI{\pm 0.4029}{\volt},\SI{0}{\volt})$ & $\lambda_1=54299,\lambda_{2,3}=-8029\pm j24043$ & unstable index-1 saddle-foci\\ \cline{2-4}
		& $P_{4\pm}:(\SI{\pm 2.1308}{\volt},\SI{\pm 0.5327}{\volt},\SI{0}{\volt})$ & $\lambda_{1,2}=1336\pm j18985,\lambda_3=-93024$ & unstable index-2 saddle-foci\\ \cline{2-4}
		& $P_{5\pm}:(\SI{\pm 2.6786}{\volt},\SI{\pm 0.6696}{\volt},\SI{0}{\volt})$ & $\lambda_1=50719,\lambda_{2,3}=-8464\pm j23863$ & unstable index-1 saddle-foci\\ \cline{2-4}
		& $P_{6\pm}:(\SI{\pm 3.1962}{\volt},\SI{\pm 0.7991}{\volt},\SI{0}{\volt})$ & $\lambda_{1,2}=1411\pm j18751,\lambda_3=-89017$ & unstable index-2 saddle-foci\\ \cline{2-4}
		& $P_{7\pm}:(\SI{\pm 3.7463}{\volt},\SI{\pm 0.9366}{\volt},\SI{0}{\volt})$ & $\lambda_1=47655,\lambda_{2,3}=-8876\pm j23655$ & unstable index-1 saddle-foci\\ \cline{2-4}
		& $P_{8\pm}:(\SI{\pm 4.2615}{\volt},\SI{\pm 1.0654}{\volt},\SI{0}{\volt})$ & $\lambda_{1,2}=1477\pm j18526,\lambda_3=-85508$ & unstable index-2 saddle-foci\\ \hline
		
		\multirow{9}{*}{9} & $P_0:(\SI{0}{\volt}, \SI{0}{\volt}, \SI{0}{\volt})$ & $\lambda_1=61005,\lambda_{2,3}=-7326\pm j24273$ & unstable index-1 saddle-focus\\ \cline{2-4}
		& $P_{1\pm}:(\pm 0.4793{\rm V},\pm 0.1198{\rm V},\SI{0}{\volt})$ & $\lambda_{1,2}=1205\pm j19362,\lambda_3=-100501$ & unstable index-2 saddle-foci\\ \cline{2-4}
		& $P_{2\pm}:(\pm 0.9402{\rm V},\pm 0.2351{\rm V},\SI{0}{\volt})$ & $\lambda_1=56878,\lambda_{2,3}=-7742\pm j24146$ & unstable index-1 saddle-foci\\ \cline{2-4}
		& $P_{3\pm}:(\pm 1.4209{\rm V},\pm 0.3552{\rm V},\SI{0}{\volt})$ & $\lambda_{1,2}=1282\pm j19144,\lambda_3=-96026$ & unstable index-2 saddle-foci\\ \cline{2-4}
		& $P_{4\pm}:(\pm 1.8804{\rm V},\pm 0.4701{\rm V},\SI{0}{\volt})$ & $\lambda_1=53342,\lambda_{2,3}=-8141\pm j24000$ & unstable index-1 saddle-foci\\ \cline{2-4}
		& $P_{5\pm}:(\pm 2.3625{\rm V},\pm 0.5906{\rm V},\SI{0}{\volt})$ & $\lambda_{1,2}=1353\pm j18933,\lambda_3=-92105$ & unstable index-2 saddle-foci\\ \cline{2-4}
		& $P_{6\pm}:(\pm 2.8204{\rm V},\pm 0.7051{\rm V},\SI{0}{\volt})$ & $\lambda_1=50285,\lambda_{2,3}=-8520\pm j23839$ & unstable index-1 saddle-foci\\ \cline{2-4}
		& $P_{7\pm}:(\pm 3.3045{\rm V},\pm 0.8261{\rm V},\SI{0}{\volt})$ & $\lambda_{1,2}=1418\pm j18728,\lambda_3=-88639$ & unstable index-2 saddle-foci\\ \cline{2-4}
		& $P_{8\pm}:(\pm 3.7612{\rm V},\pm 0.9403{\rm V},\SI{0}{\volt})$ & $\lambda_1=47615,\lambda_{2,3}=-8881\pm j23662$ & unstable index-1 saddle-foci\\ \cline{2-4}
		& $P_{9\pm}:(\pm 4.2457{\rm V},\pm 1.0614{\rm V},\SI{0}{\volt})$ & $\lambda_{1,2}=1476\pm j18529,\lambda_3=-85556$ & unstable index-2 saddle-foci\\ \hline
	\end{tabular}
	\label{table:Equilibria}
\end{table*}

\subsection{Equilibrium Point and Its Stability}
\label{subsec:Equilibria}

Set $\frac{{\d}V_1}{{\d}t}$ as zero, it is easy to obtain
\begin{equation}
V_2=V_1+Rh(V_1).
\label{eq:h1}
\end{equation}
If $\frac{{\d}V_2}{{\d}t}$ and $\frac{{\d}V_3}{{\d}t}$ are both equal to
zero, one can obtain $V_{\rm 3}=0$, and
\begin{equation}
V_2 = R_3V_1/(R + R_3).
\label{eq:h2}
\end{equation}
Thus the solutions of $V_{\rm 1}$ and $V_{\rm 2}$ are the intersection points of the two curves measured by Eqs.~(\ref{eq:h1}) and (\ref{eq:h2}).

The linear element parameters of the circuit shown in Fig.~\ref{fig:Circuitschematic} are configured as follows: $R=\SI{1.5}{\kohm}$, $R_1=\SI{3}{\kohm}$, $R_2=\SI{6.6}{\kohm}$, $R_3=R_4=\SI{500}{\ohm}$, $C_1=\SI{6.8}{\nano\farad}$, and $C_2=C_3=\SI{100}{\nano\farad}$. The parameters listed in Table~\ref{table:parameters} are utilized for the 8-$N_{\rm RI}$ and 9-$N_{\rm RI}$-based Chua's diodes. Equations~(\ref{eq:h1}) and (\ref{eq:h2}) are simultaneously plotted in Fig.~\ref{fig:Equilibriumpoints}(a) and (b), respectively, where Eq.~(\ref{eq:h1}) is indicated by solid line and marked as $h_1$, and Eq.~(\ref{eq:h2}) is denoted with dash line and marked as $h_2$. As there are ($2M+$1) intersection points generated from the inductor-free Chua's circuit with $M$-$N_{\rm RI}$ based Chua's diode, all equilibrium points can be calculated and listed in Table~\ref{table:Equilibria}.

\begin{figure*}[!htb]
	\centering
	\begin{minipage}{\OneImW}
		\centering
		\includegraphics[width=\OneImW]{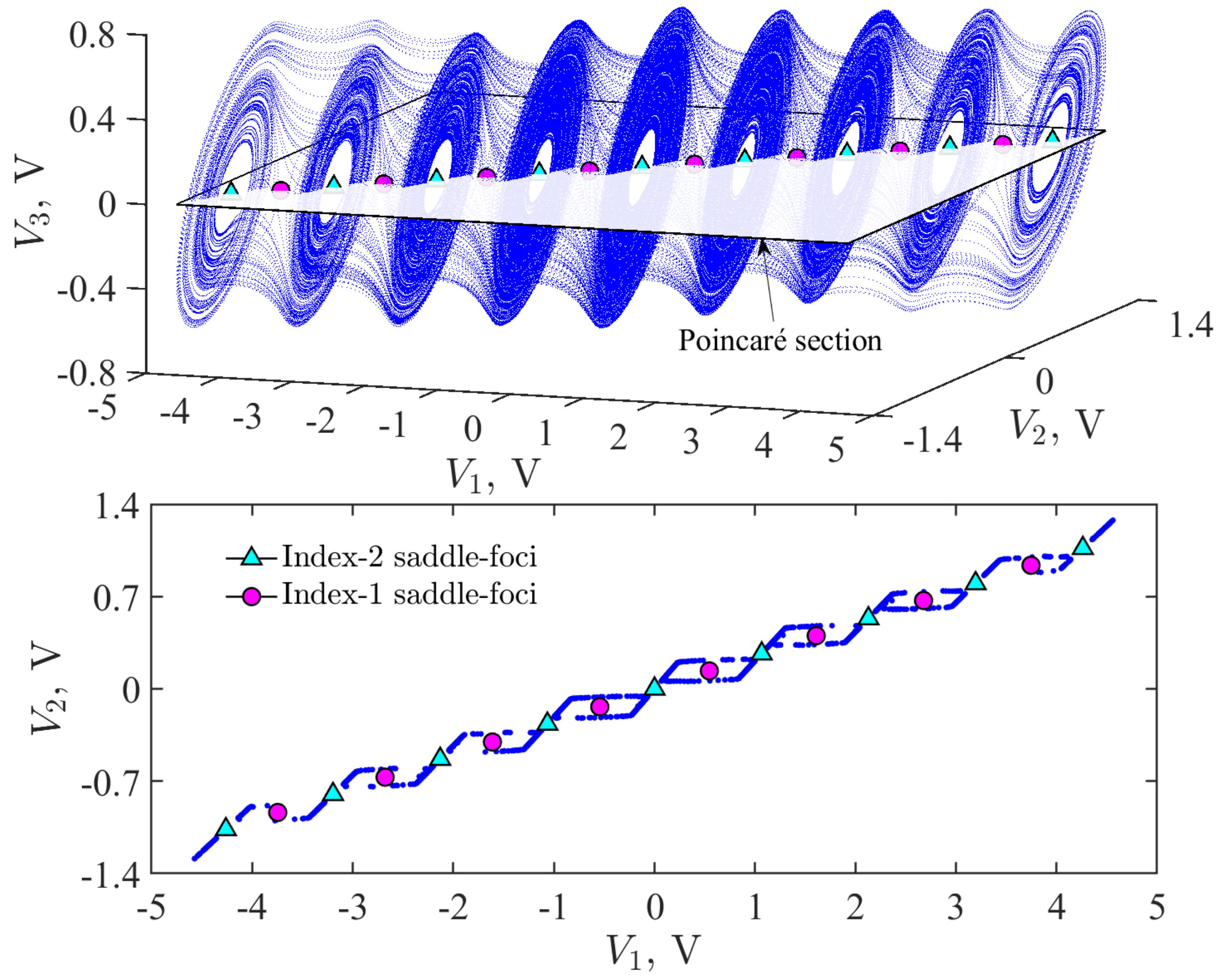}
		(a)
	\end{minipage} \hspace{5mm}
	\begin{minipage}{\OneImW}
		\centering
		\includegraphics[width=\OneImW]{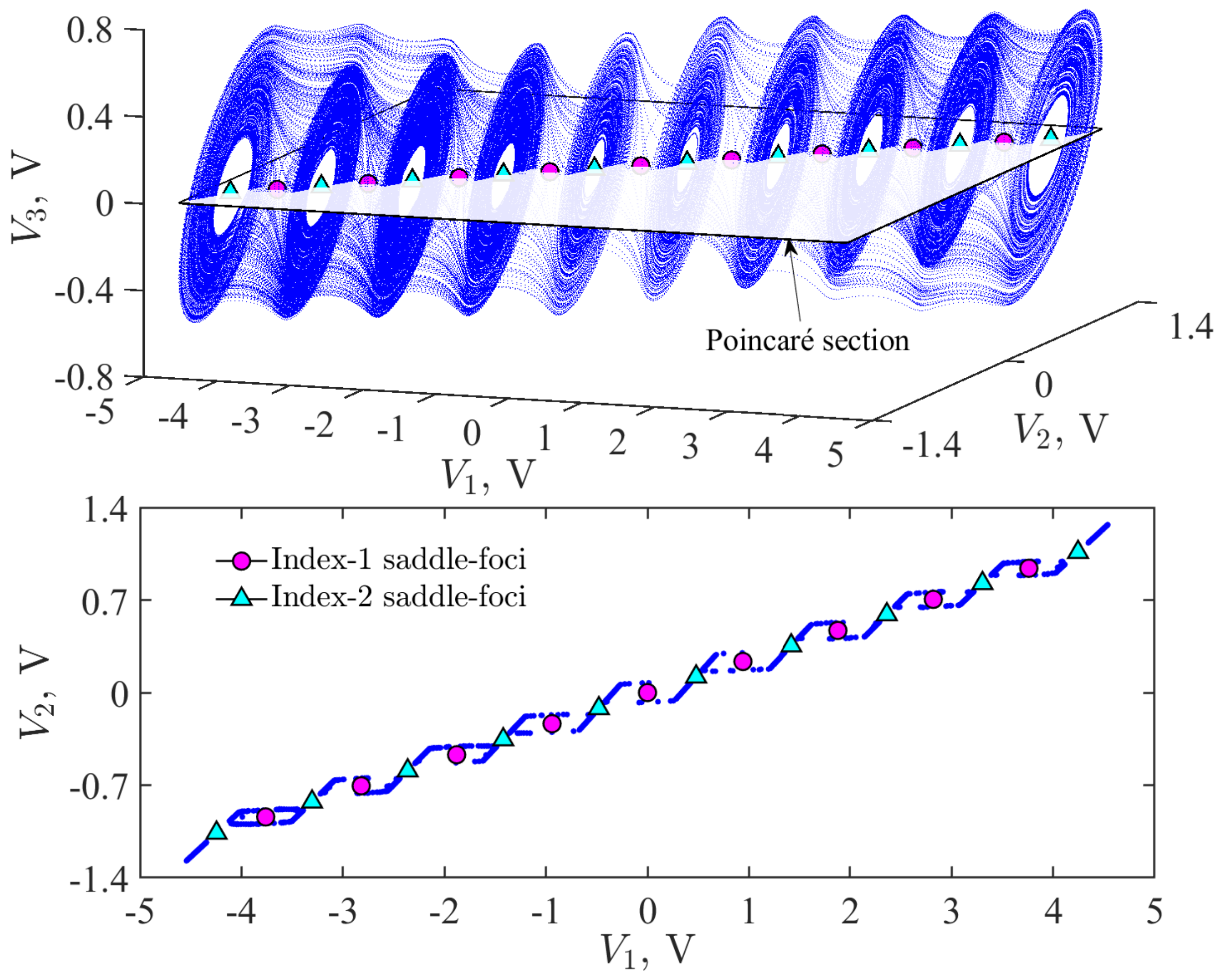}
		(b)
	\end{minipage}
	\caption{Numerical simulation results of chaotic attractors:
	(a) the Poincar\'{e} section (upper panel) and  the crossing Poincar\'{e} section of
	9-scroll chaotic attractor (lower panel); (b) the Poincar\'{e} section (upper panel) and  the crossing Poincar\'{e} section of 10-scroll chaotic attractor (lower panel).	
}
\label{fig:simulatedattractors}
\end{figure*}

The Jacobian matrix corresponding to every equilibrium point is
\begin{equation}
{\bf{J}}=\left[\begin{IEEEeqnarraybox}[][c]{,c/c/c,}
 -{\frac{1}{RC_1}-\frac{H(V_1)}{C_1}}&{\frac{1}{RC_1}}&0\\
    \frac{1}{RC_2}&-\frac{R + R_3}{RR_3C_2}&\frac{k-1}{kR_3C_2}\\
    \frac{k}{RC_3}&-\frac{k(R + R_3)}{RR_3C_3}&\frac{k-2}{R_3C_3}%
\end{IEEEeqnarraybox}\right],
\end{equation}
where
\begin{multline*}
H(V_1)=\sum\limits_{m=1}^M (-1)^m (G_{bm}+0.5(G_{am}-G_{bm})\\
(\sign(V_1+B_{pm})-\sign(V_1-B_{pm})) ).
\end{multline*}
As for the Jacobian matrices at the equilibrium points, the corresponding eigenvalues are listed in Table~\ref{table:Equilibria}, including two classes
of equilibrium points: 1) unstable index-1 saddle-foci owning a positive real root and two complex conjugate roots with negative real parts; 2) unstable index-2 saddle-foci owning two complex conjugate roots with positive real parts and a negative real root. Consequently, as for the inductor-free Chua's circuit with 8-$N_{\rm RI}$ based Chua's diode ($M=8$), there exist eight unstable index-1 saddle-foci, $P_{\rm 1\pm}$, $P_{\rm 3\pm}$, $P_{\rm 5\pm}$, $P_{\rm 7\pm}$, and nine unstable index-2 saddle-foci, $P_{\rm 0}$, $P_{\rm 2\pm}$, $P_{\rm 4\pm}$, $P_{\rm 6\pm}$, $P_{\rm 8\pm}$. As for the inductor-free Chua's circuit with 9-$N_{\rm RI}$ based Chua's diode ($M=9$), there exist nine unstable index-1 saddle-foci, $P_{\rm 0}$, $P_{\rm 2\pm}$, $P_{\rm 4\pm}$, $P_{\rm 6\pm}$, $P_{\rm 8\pm}$, and ten unstable index-2 saddle-foci, $P_{\rm 1\pm}$, $P_{\rm 3\pm}$, $P_{\rm 5\pm}$, $P_{\rm 7\pm}$, $P_{\rm 9\pm}$.

\begin{figure}[!htb]
\centering
    \begin{minipage}{\OneImW}
		\centering
        \includegraphics[width=\OneImW]{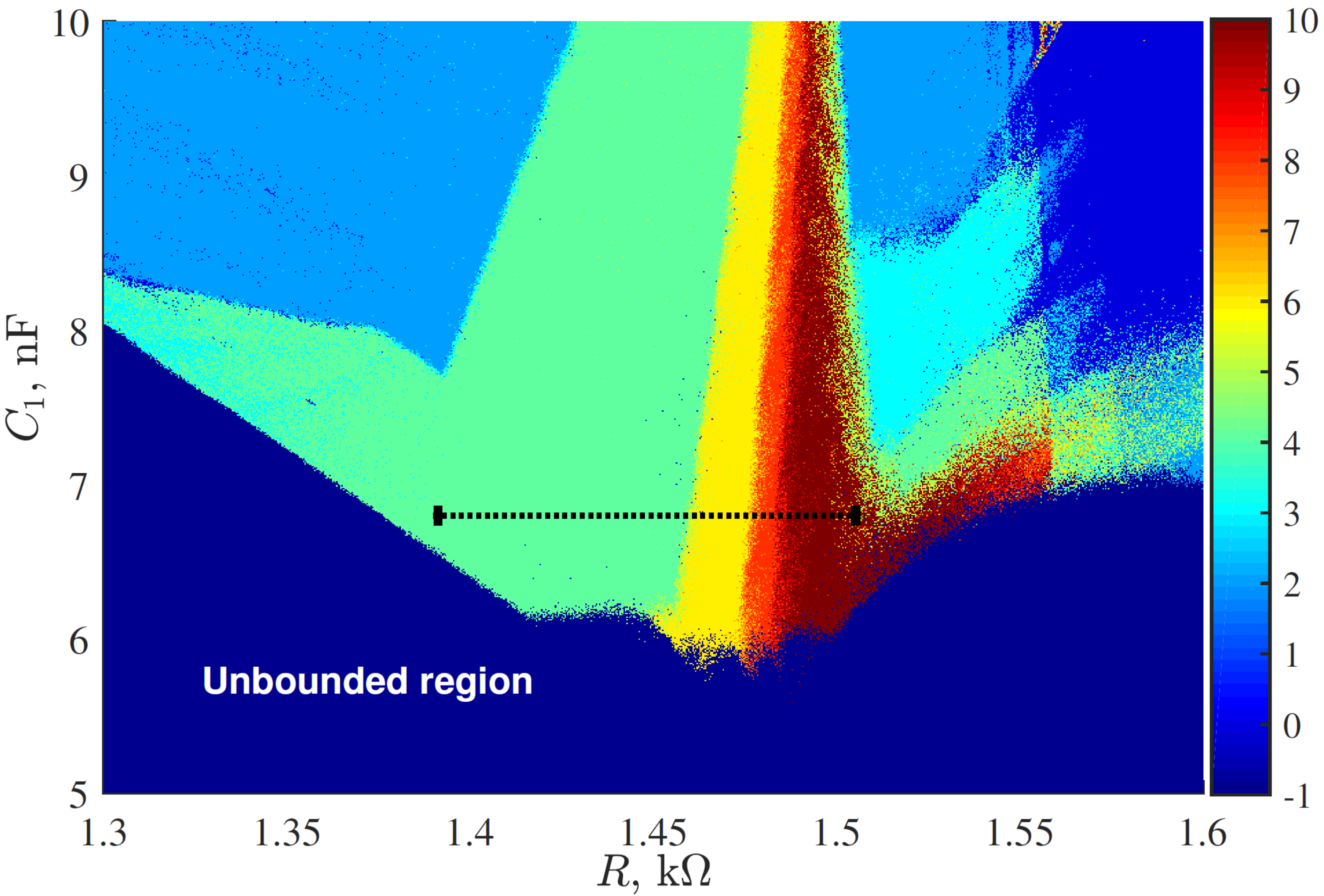}
		(a)
	\end{minipage}\\
    \begin{minipage}{\OneImW}
		\centering
        \includegraphics[width=\OneImW]{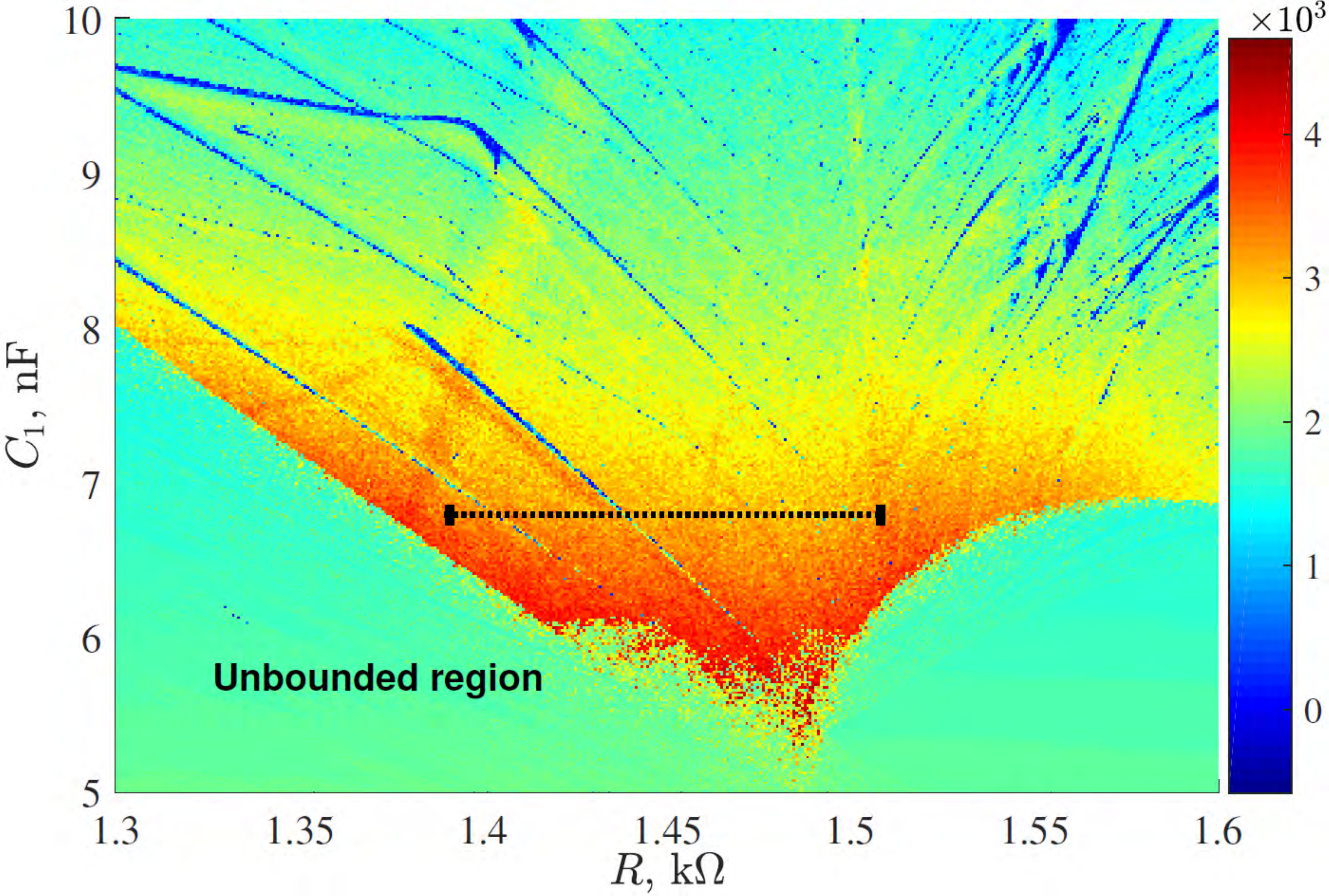}
		(b)
	\end{minipage}\\
\caption{The dynamical behaviors with respect to $R$ and $C_1$, where the black dot-line is a case of $C_1= \SI{6.8}{\nano\farad}$ and $R \in[\SI{1.385}{\kohm}, \SI{1.515}{\kohm}]$: (a) the dynamics map for detecting scrolls; (b) the largest Lyapunov exponent map.}
\label{fig:RC1map}
\end{figure}

\begin{figure}[!htb]
\centering
    \begin{minipage}{\OneImW}
		\centering
        \includegraphics[width=\OneImW]{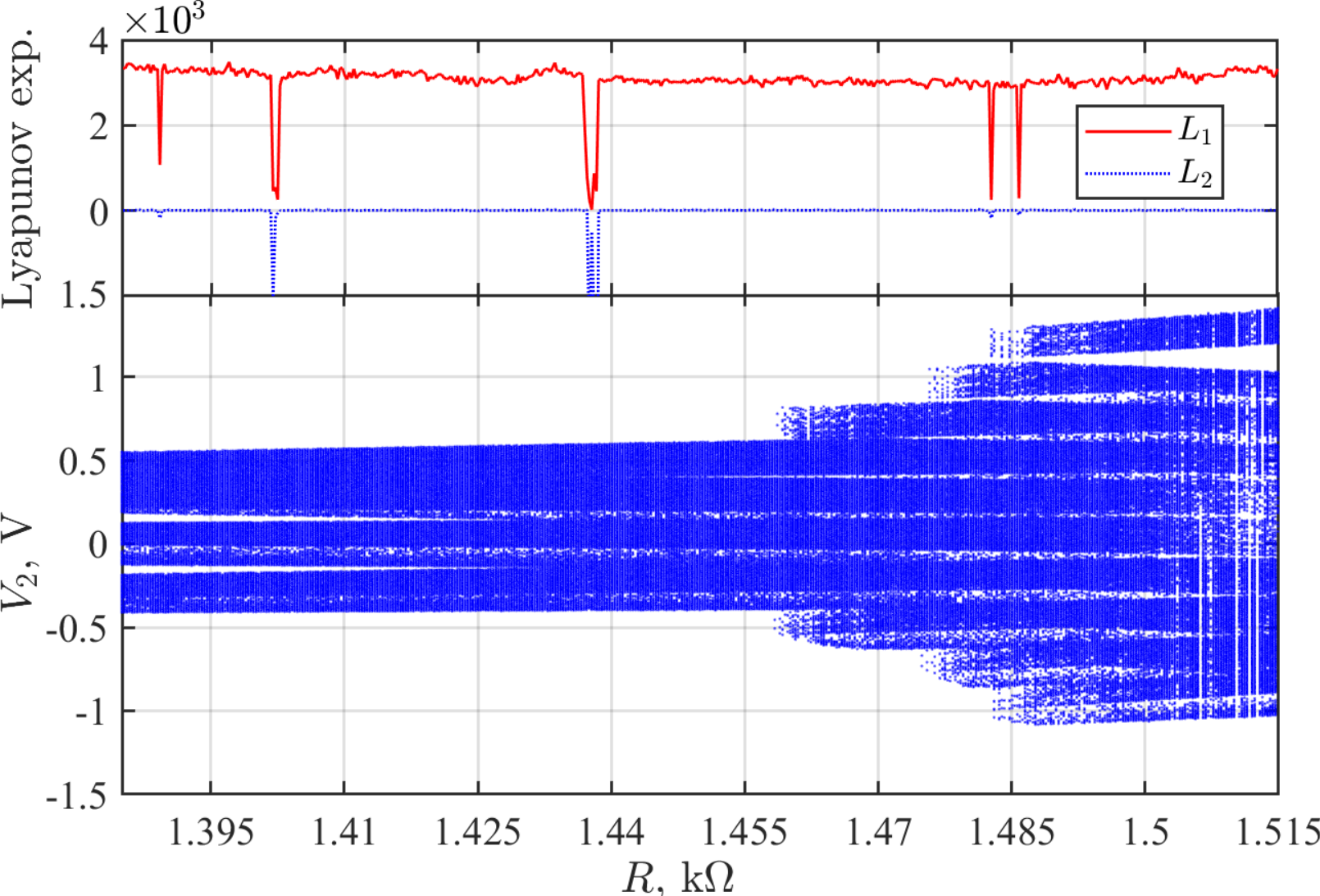}
		(a)
	\end{minipage}\\
	\begin{minipage}{\OneImW}
		\centering
		\includegraphics[width=\OneImW]{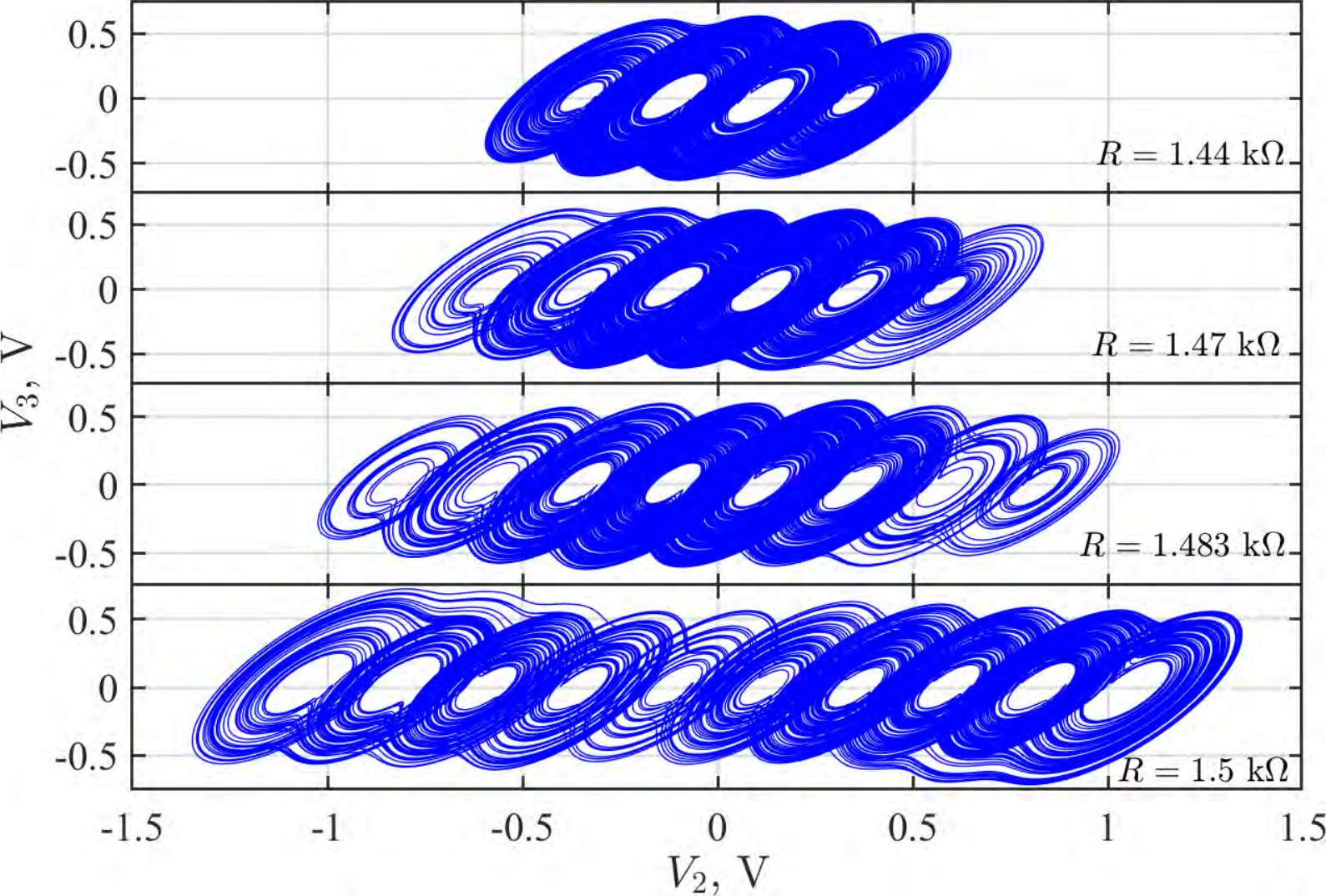}
		(b)
	\end{minipage}
\caption{The dynamical behaviors with respect to $R$: (a) the first two Lyapunov exponents (the upper panel) and the bifurcation diagram (the lower panel) of the local maxima of $V_2$; (b) the 4-, 6-, 8-, and 10-scroll chaotic attractors.}
\label{fig:Rdynamics}
\end{figure}

According to Shil'nikov theorem given in \cite{Silva1993Shil}, chaos may emerge in a system if
there is one real root $\gamma$ and two complex conjugate roots $\sigma$ $\pm$ $j \omega$ satisfying $|\frac{\sigma}{\gamma}|<1$ and $\sigma \gamma<0$ among the eigenvalues of its equilibrium points.
When suitable configuration parameters are adopted, the corresponding bond orbits and scrolls may be generated in the neighborhoods of unstable index-1 and index-2 equilibria, respectively.

\subsection{Multi-Scroll Chaotic Attractor}
\label{subsec:Attractors}

To verify the above theoretical analyses, a large number of experiments were performed using
Runge-Kutta algorithm (``ode45" in MATLAB) with time-step \SI{1}{\micro\second} and time-interval [\SI{10}{\milli\second}, \SI{200}{\milli\second}]. The obtained experimental results confirmed that no less than 20 scrolls can be generated from the proposed circuit by modifying the Chua's diode. Only two typical examples are demonstrated in Fig.~\ref{fig:simulatedattractors}.
When the initial conditions of three state variables are set as $(V_1(0), V_2(0), V_3(0)) = (\SI{0.1}{\milli\volt}, \SI{0}{\volt}, \SI{0}{\volt})$, the phase portraits and Poincar\'{e} maps of a 9-scroll attractor and a 10-scroll one are plotted in Fig.~\ref{fig:simulatedattractors}(a) and (b), respectively. It is found that the scrolls emerge from the neighbourhoods of unstable index-2 saddle-foci, while the bond orbits are generated from the neighbourhoods of unstable index-1 saddle-foci.

Generally, a Poincar\'{e} map can be used to preliminarily distinguish the state of the motion: the crossing trajectory of discrete points means the periodic state; the crossing trajectory of continuous curve means the chaotic state; the crossing trajectory of plane with no regular limbs implies the hyper-chaotic state. Considering the equilibrium points are always distributed in the $V_1-V_2$ plane, the plane of $V_3=0$ is chosen as the Poincar\'{e} section for better observation of the locations of the scrolls and the state of the whole attractor. The projections of Poincar\'{e} maps of the 9-scroll and 10-scroll chaotic attractors in the $V_1-V_2$ plane are plotted in the lower panels of Fig.~\ref{fig:simulatedattractors}(a) and (b), respectively. It can be observed that the continuous curve-like maps occur in the neighbourhoods of unstable index-2 saddle-foci, indicating the emergence of multi-scroll chaotic attractors.

In theory, the proposed scheme can construct ($2M+1$)-segment piecewise linear curve by adjusting the $3M$ parameters of the $M$-$N_{\rm RI}$-based Chua's diode. When suitable parameters are configured, the produced ($M+1$)-scroll attractor
is coined with that generated by the Chua's circuit.

\subsection{Parameter-Dependent Bifurcation Behavior}
\label{subsec:Dynamics}

Like most chaotic systems, control parameter is one of the most important factors determining system dynamics. To ensure the breakpoints and slopes are regular, the element parameters of Chua's diode are configured by the method proposed in Sec.~\ref{subsec:ChuaDiode}, and kept unchanged. The Sallen-Key HPF with determined parameters is used as an oscillating cell like the LC network of classical Chua's circuit \cite{Fortuna2009Chua}. According to Eqs.~(\ref{eq:h1}) and (\ref{eq:h2}), it is found that the equilibrium points can be changed by adjusting resistance $R$. Besides, the oscillating frequency of the active $N_RC_1$ network is determined by capacitance $C_1$. Thus $R$ and $C_1$ are two important adjustable parameters to control the system's dynamical behaviors.

\begin{figure}[!htb]
\centering
    \begin{minipage}{\OneImW}
		\centering
        \includegraphics[width=\OneImW]{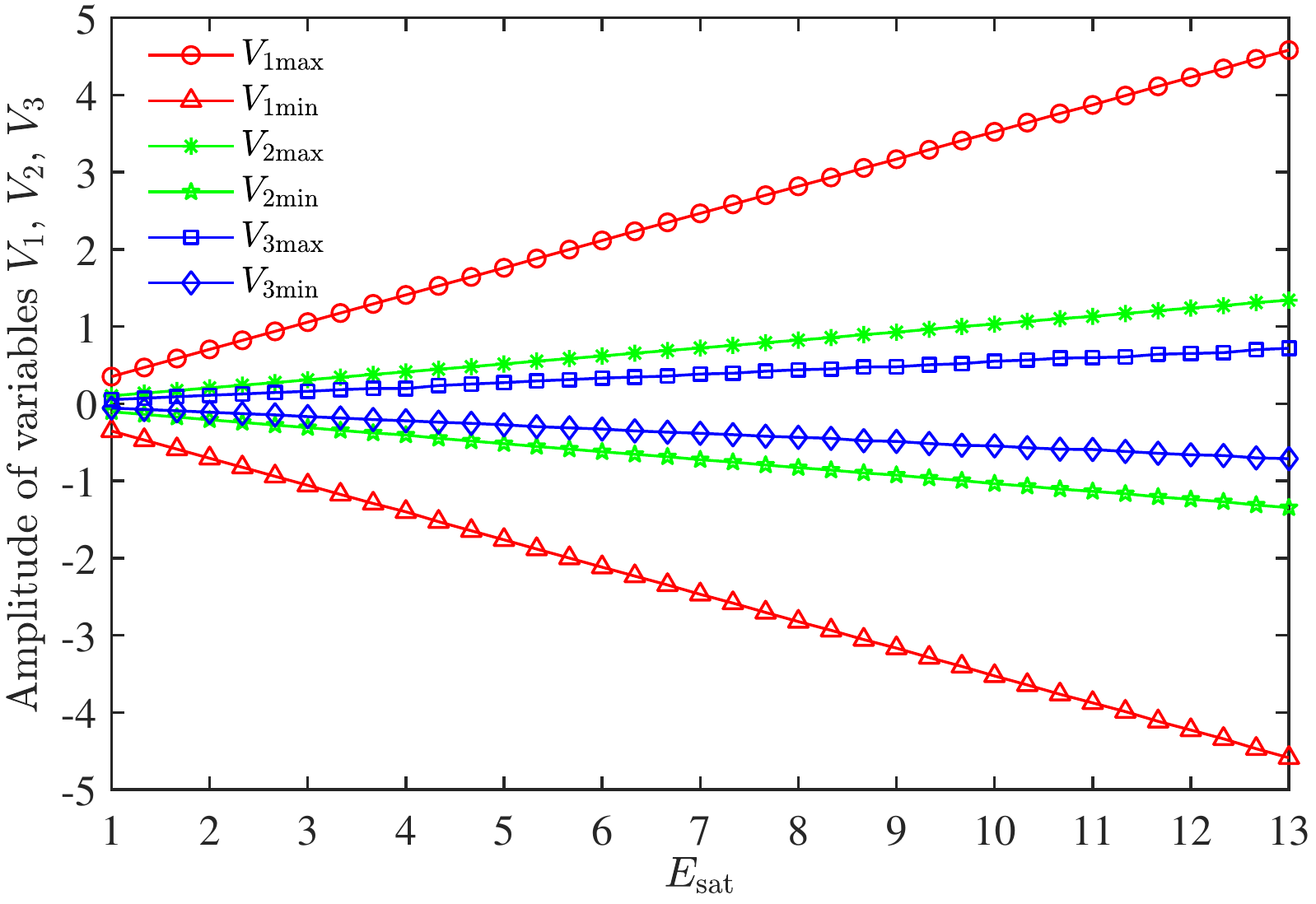}
		(a)
	\end{minipage}\\
	\begin{minipage}{\OneImW}
		\centering
		\includegraphics[width=\OneImW]{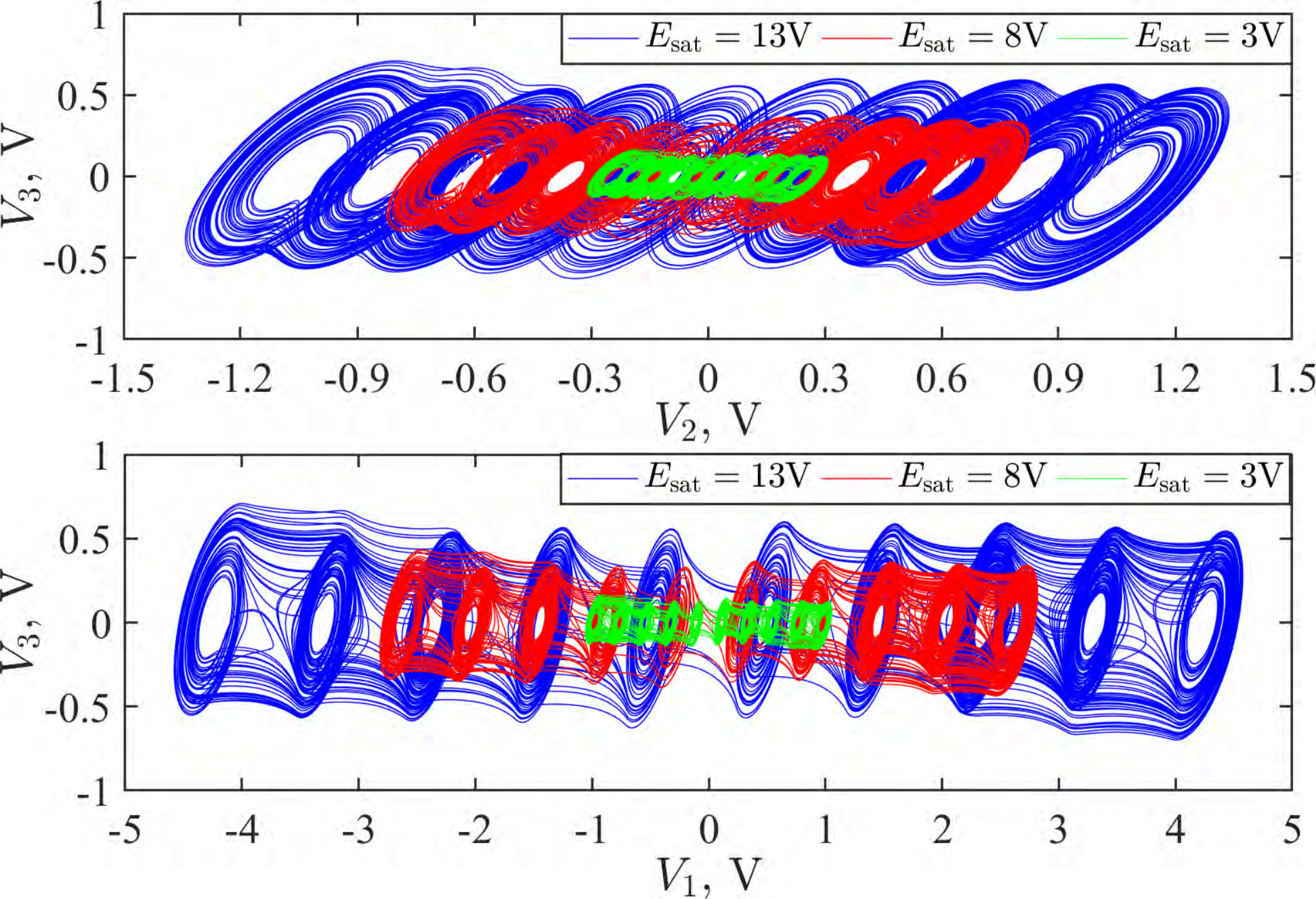}
		(b)
	\end{minipage}
\caption{The dynamical behaviors with respect to $E_{\rm sat}$: (a) the amplitude diagram of the maxima and minima of variables $V_1$, $V_2$, and $V_3$; (b) the phase portrait of 10-scroll chaotic attractors in the $V_2-V_3$ and $V_1-V_3$ planes.}
\label{fig:Edynamics}
\end{figure}

To study dynamical behaviors of the proposed system, the Sallen-Key HPF-based Chua's circuit with various Chua's diodes was measured in terms of Lyapunov exponents and bifurcation diagram.
First the parameters of Sallen-Key HPF are fixed as in Sec.~\ref{subsec:Equilibria}, and the parameters of Chua's diode with 19 segments are listed in Table~\ref{table:parameters}. To explore dynamics and recognize the scrolls, the two-parameter dynamics map and its largest Lyapunov exponent in specific region are plotted in Fig.~\ref{fig:RC1map}(a) and (b), respectively, where $R\in[\SI{1.3}{\kohm}, \SI{1.6}{\kohm}]$ and $C_1\in[\SI{5}{\nano\farad}, \SI{10}{\nano\farad}]$. Observing Fig.~\ref{fig:RC1map}, one can see that different types of dynamical regions and scrolls can be recognized by referring to the colorbar.

When capacitance $C_1$ is fixed as \SI{6.8}{\nano\farad} and the resistance $R$ is incrementally changed from \SI{1.385}{\kohm} to \SI{1.515}{\kohm}, the first two Lyapunov exponents calculated by Wolf's method given in \cite{Wolf1985Determining} and bifurcation diagram of the local maxima (denoted by $V_2$) of the state variable $V_2$ are plotted in Fig.~\ref{fig:Rdynamics}(a). It is observed from Fig.~\ref{fig:Rdynamics}(a) that there exists a wide range with positive largest Lyapunov exponents when $R\in[\SI{1.385}{\kohm}, \SI{1.515}{\kohm}]$, indicating the emergence of chaos.
The dynamical behaviors depicted by the bifurcation diagram agree well with those described by the black dot-line in the dynamics map in Fig.~\ref{fig:RC1map}.
Furthermore, 4-, 6-, 8-, and 10-scroll chaotic attractors are observed.
When resistance $R$ is selected as \SI{1.44}{\kohm}, \SI{1.47}{\kohm}, \SI{1.483}{\kohm}, and \SI{1.5}{\kohm}, the phase portraits in the $V_2-V_3$ plane are plotted in Fig.~\ref{fig:Rdynamics}(b). The results of Fig.~\ref{fig:RC1map} and \ref{fig:Rdynamics} show that complex dynamical transitions and different scroll types can be observed with variation of the two parameters.

When the element parameters are all fixed, the amplitude of the attractors can be scaled up or down on the saturation output voltage $E_{\rm sat}$ of the op-amps. According to Eq.~(\ref{eq:vi}), $E_{\rm sat}$ can control the breakpoints $B_{pm}$ according to $\frac{R_{am}E_{\rm{sat}}}{R_{bm}}$, while the slopes $G_{am}$ and $G_{bm}$ keep unchanged. Accordingly, in Fig.~\ref{fig:Equilibriumpoints}, the equilibria measured by the intersection points of the two curves $h_1$ and $h_2$ converge toward the origin with $E_{\rm sat}$ decreasing. Inversely, they diverge from the origin when $E_{\rm sat}$ increasing. The amplitude diagram of the 10-scroll attractors is plotted in Fig.~\ref{fig:Edynamics}(a). It is found that the maxima and minima of variables $V_1$, $V_2$, and $V_3$ linearly escalate with the increase of $E_{\rm sat}$. When $E_{\rm sat}$ is selected as \SI{13}{\volt}, \SI{8}{\volt}, and \SI{3}{\volt}, the phase portraits in the $V_1-V_3$ and $V_2-V_3$ planes are plotted in Fig.~\ref{fig:Edynamics}(b). The results of Fig.~\ref{fig:Edynamics} indicate that the attractors can be generated under different power supplies, which verifies the feasibility of satisfying the low-voltage low-power requirement.

\subsection{Comparison of Related Chua's Ciruits}
\label{subsec:Comparison}

\renewcommand{\arraystretch}{1.05}
\begin{table*}[!htb]
	\caption{Comparison of Related Multi-scroll Chua's Circuit.}
	\centering
    \begin{tabular}{c|c|c|c|c|c|c|c|c|c}
    \hline
    \multirow{2}{*}{Refs} &\multirow{2}{*}{Inductor-free}&\multirow{2}{*}{Circuit design} & \multirow{2}{*}{Experiment} & \multirow{2}{*}{Attractor type}
    &\multicolumn{5}{c}{Circuit element} \cr\cline{6-10}
                                                       &&&&&Inductor    &Capacitor   &Resistor    &Op-amp      &DC bias voltage \cr\hline
    \cite{Suykens1997}          &No   &No  &No  &7-scroll  &$\verb|\|$  &$\verb|\|$  &$\verb|\|$  &$\verb|\|$  &$\verb|\|$  \cr\hline
    \cite{Yal2000Experimental}  &No   &Yes &Yes &5-scroll  &1           &3           &30          &8           &$\verb|\|$  \cr\hline
    \cite{GUOQUNZHONG2002A}     &No   &Yes &Yes &10-scroll &1           &2           &31          &10          &8           \cr\hline
    \cite{Yu2003New}            &No   &Yes &Yes &10-scroll &1           &3           &54          &13          &$\verb|\|$  \cr\hline
    \cite{Karthikeyan2018NOVEL} &Yes  &Yes &No  &4-scroll  &$\verb|\|$  &3           &48          &21          &4           \cr\hline
    This work                   &Yes  &Yes &Yes &10-scroll &$\verb|\|$  &3           &25          &11          &$\verb|\|$  \cr\hline
    \end{tabular}
    \label{table:Comparison}
\end{table*}

A variety of multi-scroll Chua's attractors were reported over the past few years. Table~\ref{table:Comparison} shows the comparison between the implementations of related multi-scroll Chua's circuits. It is found that few scrolls are obtained from \cite{Yal2000Experimental,Karthikeyan2018NOVEL} but along with a relatively more use of components. Comparing with the method in \cite{GUOQUNZHONG2002A}, we use only ten op-amps to implement 19-segment Chua's diode. Furthermore, benefiting from the proposed scheme, extra 10 resistors and 8 DC bias voltages can be removed. Since complex implementation of voltage-controlled voltage source and voltage-controlled current source are cast away, both the types and numbers of circuit elements are less than that of the methods given in \cite{Yal2000Experimental,Yu2003New}. Although the implementation complexity between different design methods and hardware platforms cannot be judged by a unified standard, the proposed inductor-free Chua's circuit is a simplified implementation in terms of the number of discrete components.

\section{Application in Image Secure Communication}
\label{sec:Application}

To satisfy requirements on transmission security, different schemes (algorithms, techniques, methods) are applied to image encryption. The representative image encryption schemes published in year 2018 are reviewed in \cite{cqli:meet:JISA19}. The specific properties of chaotic systems, such as initial state sensitive dependence, unpredictability, make them widely studied and applied in the field of information security \cite{Hua2018Sine,cqli:autoblock:IEEEM18,cqli:IEAIE:IE18,Huazy:CAT:IEETC18,cqli:network:TCASI2019}. These properties are similar to the counterparts of cryptography \cite{cqli:autoblock:IEEEM18,cqli:IEAIE:IE18,Hong2018A}. In this section, an image encryption scheme based on the proposed Chua's circuits with 19- and 21-segment piecewise-linear Chua's diodes are proposed to verify its merit for image security.

\subsection{Description of encryption process}
\label{subsec:Process}

Without lose of generality, a gray-scale image $\bm{I}$ is considered as the encryption object. Then, the process of the encryption and decryption can be described as follows.
\begin{enumerate}
\item Set initial conditions and system parameters, then iterate system~(\ref{eq:circuitequation2}) with the Runge-Kutta algorithm (Function ``ode45" in MATLAB) from initial state $(x_0, y_0, z_0)$. After continuous iteration, three chaotic sequences $V_1(t)$, $V_2(t)$ and $V_3(t)$ are generated. The selected length of the chaotic sequences is same as that of the plain-image.

\item A pseudo-random sequence with amplitude $[0,255]$ is obtained by preprocessing the selected chaotic sequence $z$ by
\begin{equation}
n(i, j)=\lfloor (z_i+|z_{\min}|)\cdot 10^7 \rfloor \bmod 256,
\label{eq:PRS}
\end{equation}
where $z_{\rm{min}}$ is the minimum value of $z_i$, $\lfloor x \rfloor$ rounds the elements of $x$ to the nearest integers less than or equal to $x$.

\item Use XOR (exclusive OR) operation
\[
\bm{I}'(i, j)=\bm{I}(i, j)\oplus n(i,j)
\]
to encrypt the original image bit-by-bit (stream cipher) until all the elements are encrypted.

\item Decryption is the reverse version of the encryption process.
\end{enumerate}

\subsection{Simulation Results and Performance Analyses}

To verify real performance of the above image encryption scheme, a number of experiments
were performed by Matlab R2017b. Furthermore, it was measured by various classic metrics:
randomness test; histogram; correlation of adjacent pixels, and information entropy.

\subsubsection{Randomness Test of Pseudorandom Number Sequence}

\renewcommand{\arraystretch}{1.2}
\begin{table*}[!htb]
	\caption{The randomness rest results on generated sequences using NIST suite.}
	\centering
    \begin{threeparttable}[b]
    \begin{tabular}{l|l|c|c|c|c|c|c|c|c}
    \hline
    \multirow{3}{*}{No.} &\multirow{3}{*}{Statistical tests}
                               &\multicolumn{4}{c|}{Sequences generated by 10-scroll chaotic attractor}
                               &\multicolumn{4}{c}{Sequences generated by 11-scroll chaotic attractor}                                         \cr\cline{3-10}
                              && \multicolumn{2}{c|}{100 samples ($n=100$)}& \multicolumn{2}{c|}{1000 samples ($n=1000$)}
                               & \multicolumn{2}{c|}{100 samples ($n=100$)}& \multicolumn{2}{c}{1000 samples ($n=1000$)}                                      \cr\cline{3-10}
                              &&Proportion\tnote{*} & \textit{U}-value & Proportion\tnote{*} & \textit{U}-value
                               &Proportion\tnote{*} & \textit{U}-value & Proportion\tnote{*} & \textit{U}-value                                \cr\hline\hline
	1&	Frequency                             & 1.00          & 0.678686  & 0.995  & 0.444619  & 0.99          & 0.026948  & 0.989  & 0.550347 \cr\hline
	2&	Block frequency                       & 0.98          & 0.883171  & 0.992  & 0.408275  & 0.99          & 0.366918  & 0.990  & 0.174728 \cr\hline
	    \multirow{2}{*}{3}
     &	Cumulative sums (forward)             & 1.00          & 0.383827  & 0.993  & 0.370262  & 1.00          & 0.102526  & 0.992  & 0.645448 \cr\cline{2-10}
     &  Cumulative sums (reverse)             & 0.99          & 0.867692  & 0.990  & 0.783019  & 1.00          & 0.574903  & 0.989  & 0.801865 \cr\hline
	4&	Runs                                  & 0.97          & 0.699313  & 0.991  & 0.910091  & 0.99          & 0.798139  & 0.990  & 0.942198 \cr\hline
    5&  Longest runs                          & 1.00          & 0.678686  & 0.990  & 0.977480  & 0.99          & 0.334538  & 0.994  & 0.500279 \cr\hline
	6&	Rank                                  & 0.99          & 0.574903  & 0.991  & 0.002906  & 0.99          & 0.304126  & 0.990  & 0.308561 \cr\hline
    7&  FFT                                   & 0.98          & 0.955835  & 0.987  & 0.236810  & 0.98          & 0.574903  & 0.987  & 0.554420 \cr\hline
	8&	Nonoverlapping template\tnote{\dag}   & 0.97          & 0.574903  & 0.981  & 0.518106  &\textbf{0.95}  & 0.867692  & 0.982  & 0.363593 \cr\hline
    9&  Overlapping template                  & 0.99          & 0.304126  & 0.990  & 0.307077  & 0.98          & 0.262249  & 0.985  & 0.701366 \cr\hline
	10&	Universal                             & 1.00          & 0.534146  & 0.989  & 0.251837  & 0.99          & 0.236810  & 0.984  & 0.088226 \cr\hline
    11& Approximate entropy                   & 0.98          & 0.383827  & 0.986  & 0.010457  & 0.97          & 0.514124  & 0.991  & 0.548314 \cr\hline
	12&	Random excursions\tnote{\dag}         & 0.9636        & 0.224821  & 0.9821 & 0.432747  &\textbf{0.9524}& 0.364146  & 0.9820 & 0.144921 \cr\hline
    13& Random excursions variant\tnote{\dag} &\textbf{0.9272}& 0.366918  & 0.9837 & 0.134475  & 0.9841        & 0.037157  & 0.9820 & 0.309707 \cr\hline
	\multirow{2}{*}{14}
      & Serial (1st sub-test)                 & 1.00          & 0.595549  & 0.981  & 0.564639  & 0.99          & 0.883171  & 0.992  & 0.235589 \cr\cline{2-10}
      & Serial (2nd sub-test)                 & 1.00          & 0.834308  & 0.987  & 0.114712  & 0.99          & 0.419021  & 0.993  & 0.765632 \cr\hline
    15& Linear complexity                     & 0.97          & 0.834308  & 0.988  & 0.719747  & 0.99          & 0.350485  & 0.989  & 0.906069 \cr\hline
    \end{tabular}
    \begin{tablenotes}
    \item[*] \textit{The proportion of sequences that pass is computed by the empirical results for a particular statistical test.
                     For passing the test, the proportion value is 0.96 for 100 samples, and 0.98 for 1000 samples, respectively.}
     \item[\dag] \textit{The Nonoverlapping template test, Random excursions test, and Random excursions variant test consist of 148, 8, and 18 sub-tests, respectively.
                      As for test item composing of more than one sub-tests, the worst result is reported.}
    \end{tablenotes}
    \end{threeparttable}
    \label{table:NISTtest}
\end{table*}

The pseudorandom number generator designed in the above subsection was test by the NIST-800-22 statistical test suite with the default significance level. The initial state $(x_0, y_0, z_0)$ are randomly configured within interval $(0, 0.6)$. When time length \SI{125}{\milli\second} is used, a pseudorandom sequence of length $\frac{\SI{125}{\milli\second}}{\SI{1}{\micro\second}}=1.25\cdot10^5$ is generated by Eq.~(\ref{eq:PRS}). The corresponding binary sequence of length $1.25 \cdot10^5\cdot 8=10^6$ is obtained via converting the integer pseudorandom number to the binary format. Finally, in the file \textit{finalAnalysisReport.txt}, the interval between 0 and 1 is divided into 10 sub-intervals, and the \textit{P}-values falling in each sub-interval $C_1, C_2, \cdots, C_{10}$ are counted. The corresponding \textit{U}-values can be calculated to measure uniform distribution of the \textit{P}-values. If \textit{U}-values are larger than or equal to 0.0001, the degree of uniform distribution can be considered
efficiently high. Note that at least 55 bitstreams should be processed to provide statistically meaningful results.
As \cite{Panda:PRBG:TCASI2019}, the results on 15 tests items for 100 and 1000 bitstreams generated with 10-scroll and 11-scroll chaotic attractors are given in Table~\ref{table:NISTtest}, demonstrating that \textit{U}-Values are all larger than 0.0001. Note that two among 18 sub-tests fail to meet the satisfying ratio a little when 100 bitstreams generated with 10-scroll chaotic attractor are tested. 
As for the 100 bitstreams generated with 11-scroll chaotic attractor, the passing proportion below the expected threshold with a marginal value.
When the sample number of sequences is increased to 1,000, all test items can be passed
well for the two sources of PRNS. The confidence interval is reasonably accurate for large sample sizes (e.g., $n\geq1000$). So the designed pseudorandom number generator has good randomness and is suitable for image encryption application.

\begin{figure}[!htb]
\centering
\includegraphics[width=1.2\OneImW]{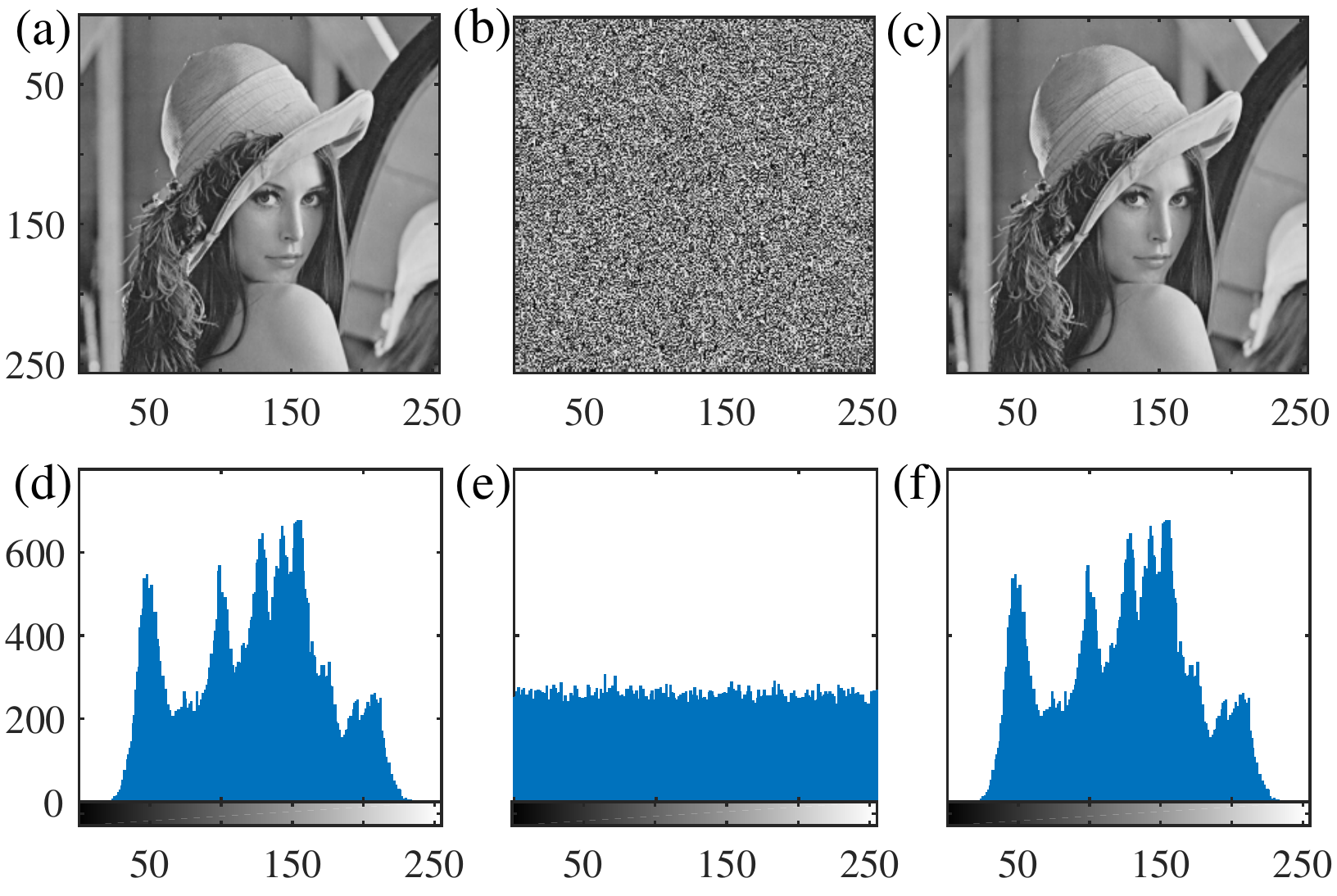}
\caption{Simulation results of the proposed encryption and decryption scheme: (a) original image; (b) cipher-image; (c) decrypted image; (d) histogram of the original image; (e) histogram of the cipher-image; (f) histogram of the decrypted image.}
\label{fig:IamgeSimulation}
\end{figure}

\begin{figure}[!htb]
\centering
\includegraphics[width=1.2\OneImW]{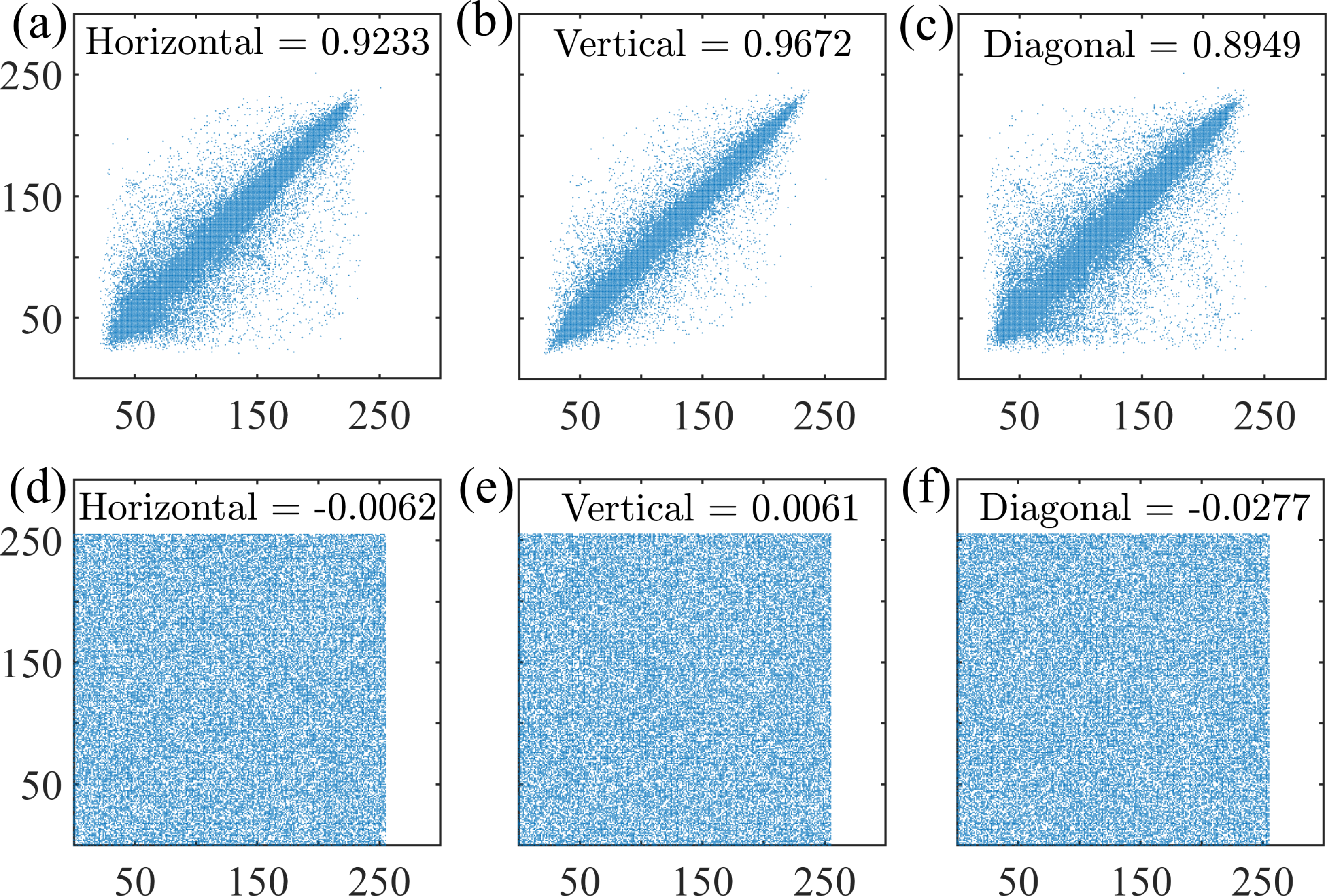}
\caption{Correlation and correlation coefficients of adjacent pixels image ``Lenna" and its cipher-image: (a) horizontal correlation of the original image; (b) vertical correlation of the original image; (c) diagonal correlation of the original image; (d) horizontal correlation of the cipher-image; (e) vertical correlation of the cipher-image; (f) diagonal correlation of cipher-image.}
\label{fig:Correlation}
\end{figure}

\subsubsection{Histogram Analysis}	

Every plain-image has a histogram measuring unique distribution regularity
on pixel intensity values. A good encryption process can erase strong correlation among neighbouring pixels of the plain-image, and make the histogram fluctuation as small as possible \cite{cqli:IEAIE:IE18}. The initial state $(x_0, y_0, z_0)$ is set  as $(0.1,0.1,0.1)$, and a pseudorandom sequence generated by Eq.~(\ref{eq:PRS}) is used for image encryption.  Figure~\ref{fig:IamgeSimulation}(a), (b), and (c) depict the plain-image, cipher-image, and decrypted image, respectively.
In contrast, the corresponding histograms are displayed in Fig.~\ref{fig:IamgeSimulation}(d), (e), and (f). It is found that the histogram of the cipher-image is very uniform, which means that the encryption scheme based on the proposed chaotic system has a good performance on robustness withstanding the statistical attacks.

\subsubsection{Correlation Analysis}

The adjacent pixels of a natural image have a strong correlation in the horizontal, vertical, and diagonal directions. But in a cipher-image, the correlation coefficients are expected to be close to zero in three directions. The correlation of each pair of pixels can be calculated by
\[C = \frac{\sum\nolimits_{i=1}^N{({x_i}-{\textstyle{1 \over N}}\sum\nolimits_{i=1}^N {x_i})({y_i}-{\textstyle{1 \over N}}\sum\nolimits_{i = 1}^N {{y_i}} )} }{\sqrt {\sum\nolimits_{i = 1}^N {{({x_i} - {\textstyle{1 \over N}}\sum\nolimits_{i = 1}^N {{x_i}} )}^2} } \sqrt{\sum\nolimits_{i=1}^N {{{({y_i} - {\textstyle{1 \over N}}\sum\nolimits_{i = 1}^N {{y_i}} )}^2}} } }\]
where $x$ and $y$ are the intensity values of two adjacent pixels, and $N$ is the total number of pixels.

We randomly select 1,000 pairs of adjacent pixels from the plain image and cipher image to test their internal correlations. The adjacent pixels correlation and correlation coefficients of the plain and cipher images are given in Fig.~\ref{fig:Correlation}. The results show that the plain image has high correlation values, while the cipher image has much lower ones. It is indicated that the proposed encryption scheme can effectively reduce the correlation between adjacent pixels of the encrypted image.

\subsubsection{Information Entropy}

The information entropy is an important indicator in evaluating the randomness of a cipher-image. If an encryption scheme can generate cipher-images owning the maximum information entropy close to eight, it means that it has excellent randomness property. The Shannon entropy of image $\bm{I}$ is defined by
\[
H(\bm{I})=-\sum\nolimits_{i=1}^L \Pr(\bm{I}_i) \log_2\Pr(\bm{I}_i),
\]
where $\bm{I}_i$ denotes the $i$-th possible value of $\bm{I}$, ${\rm Pr}(\bm{I}_i)$ represents the probability value of $\bm{I}_i$, and $L$ is the number of possible different intensities. The calculated entropy value of the cipher-image is 7.9889 and close to the ideal value. Besides, the pseudo-random sequences generated from 11-scroll chaotic attractor are used to encrypt the same plain-image. The information entropy of the obtained cipher-image is 7.9898. It is found that there is a slight improvement in performance of cryptosystems when increasing one scroll.

All the above analysis demonstrate that the image encryption scheme based on the proposed chaotic system has good security performance and can be used for protecting image data in cyberspace.

\section{Ciruit Simulations and Hardware Experiments}
\label{sec:Experiments}

Circuit simulations and hardware experiments are necessary for circuit syntheses and verifications. Some regular circuit design and simulation software, for examples, Pspice \cite{Rocha2009An,maj:scro:ND14,Hong2017A,Hong2018A}, Multisim \cite{Huang2015Novel, Wang2018Emerging,Bao2016Inductor}, and PSIM \cite{Bao2018Initial} can provide reliable operating environment for circuit simulations. In this section, the simulation results of the obtained system by multisim circuit simulations and hardware experiments were described to verify its real performance.

\subsection{Multisim Circuit Simulations}
\label{subsec:Multisim}

\begin{figure}[!htb]
	\centering
	\begin{minipage}{\OneImW}
		\centering
		\includegraphics[width=\OneImW]{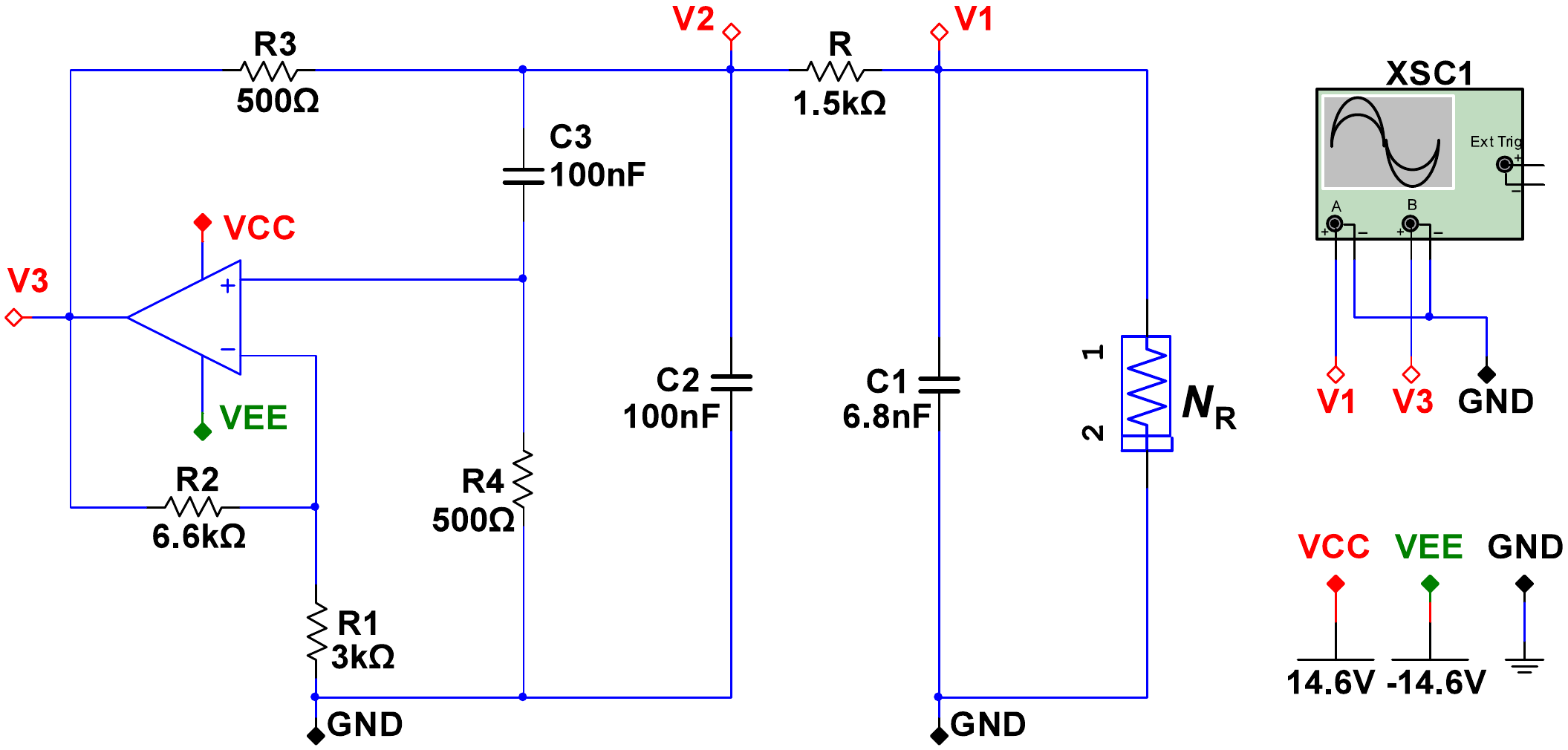}
		(a)
	\end{minipage}\\
	\begin{minipage}{\OneImW}
		\centering
		\includegraphics[width=\OneImW]{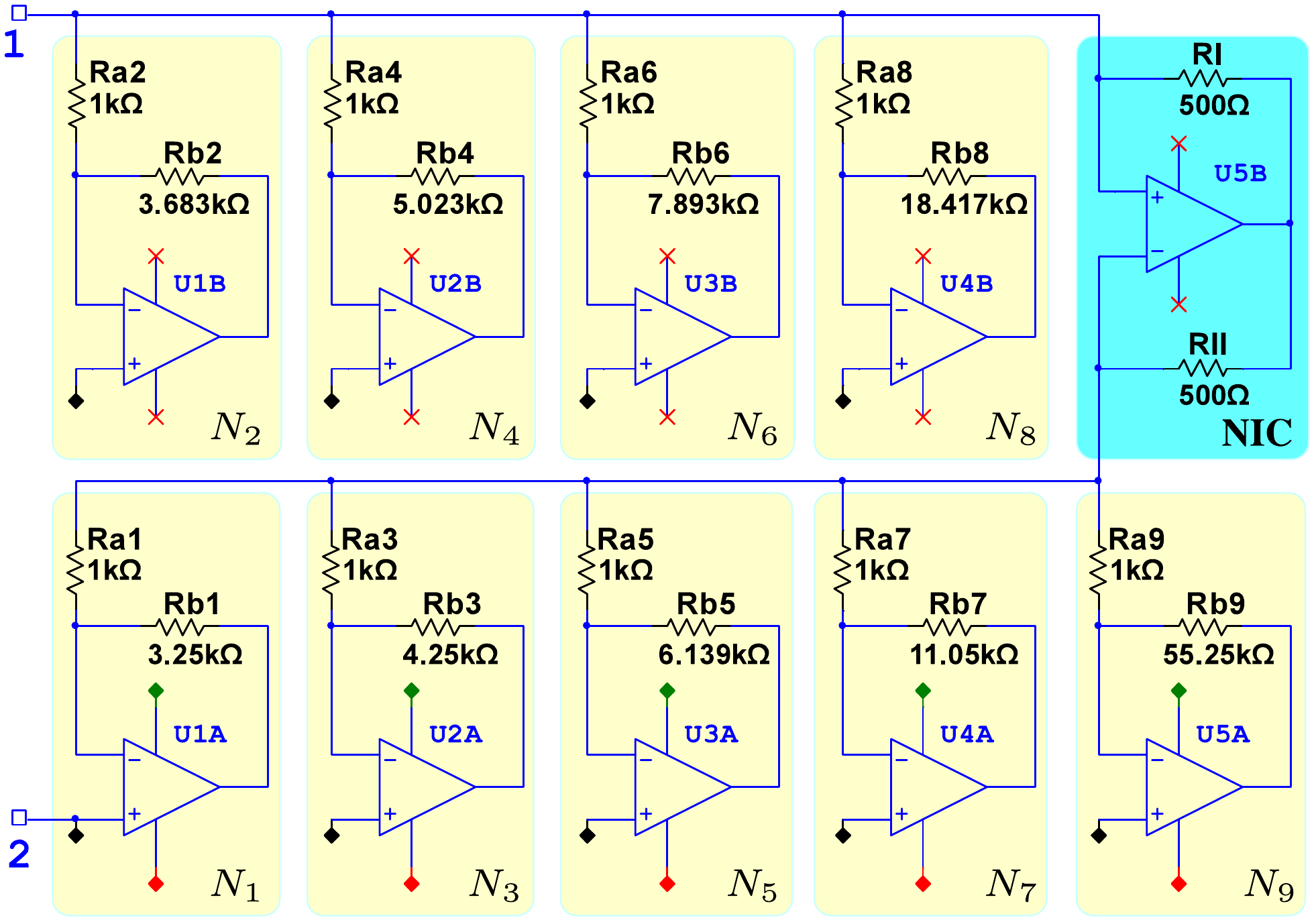}
		(b)
	\end{minipage}
	\caption{Multisim circuit simulation model: (a) the Sallen-Key HPF-based inductor-free Chua's circuit; (b) the implementation of 9-$N_{\rm RI}$ based Chua's diode.}
	\label{fig:Multisimcircuit}
\end{figure}

\begin{figure}[!htb]
	\centering
	\begin{minipage}{\OneImW}
		\centering
		\includegraphics[width=\OneImW]{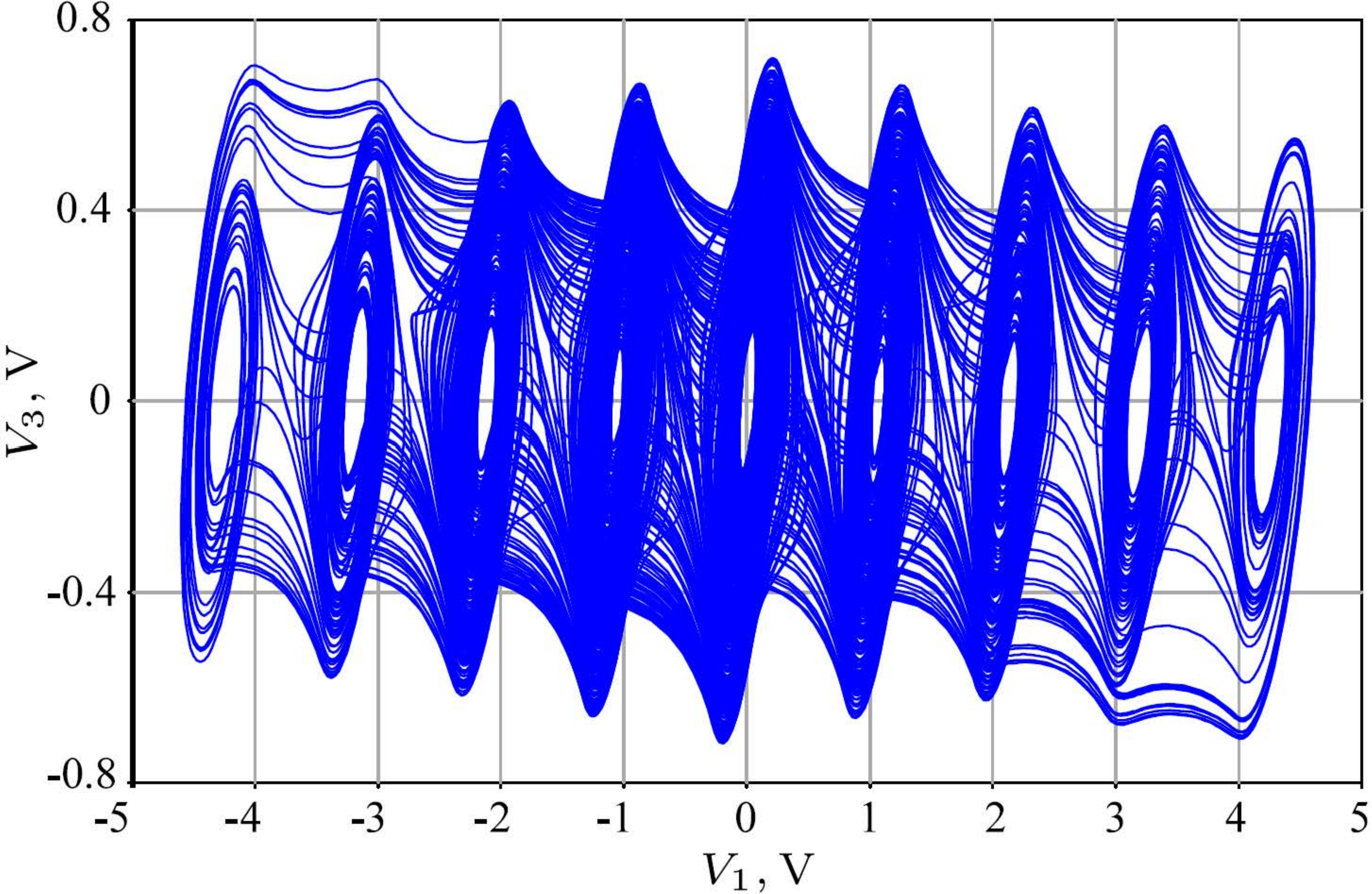}
		(a)
	\end{minipage}\\
	\begin{minipage}{\OneImW}
		\centering
		\includegraphics[width=\OneImW]{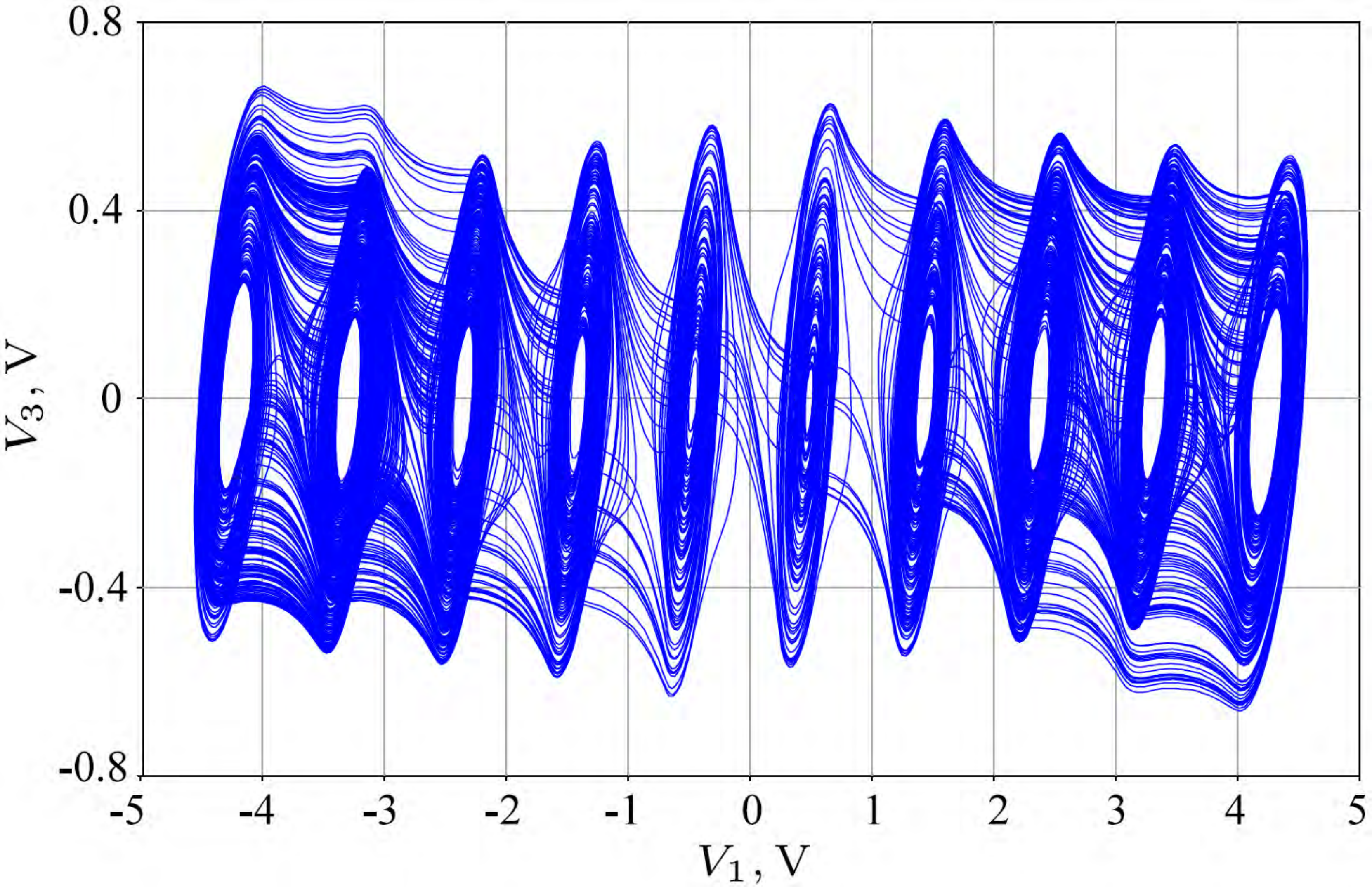}
		(b)
	\end{minipage}
	\caption{Chaotic attractors simulated in circuit: (a) the 9-scroll chaotic attractor; (b) the 10-scroll chaotic attractor.}
	\label{fig:Multisimsimulatedattractors}
\end{figure}

Using the circuit schematics shown in Figs.~\ref{fig:Chuadiode} and~\ref{fig:Circuitschematic}, the circuit simulation model of the Sallen-Key HPF-based Chua's circuit can be constructed using Multisim 12.0 simulation software. As a typical example, the Chua's circuit with 9-$N_{\rm RI}$-based Chua's diode is given in Fig.~\ref{fig:Multisimcircuit}. Using the parameters of Chua's diode listed in Table~\ref{table:parameters} and the element parameters given in Sec.~\ref{subsec:Equilibria}, the phase portraits of the 9-scroll and 10-scroll Chua's chaotic attractors in the $V_1-V_3$ plane are shown in Fig.~\ref{fig:Multisimsimulatedattractors}(a) and (b), respectively. It is observed that the simulation results with the multisim circuit shown in Fig.~\ref{fig:Multisimsimulatedattractors} agree well that obtained with the numerical simulation results demonstrated in Fig.~\ref{fig:simulatedattractors}.

\subsection{Hardware Experiments}
\label{subsec:Hardware}

\begin{figure}[!htb]
\centering
\includegraphics[width=0.8\OneImW]{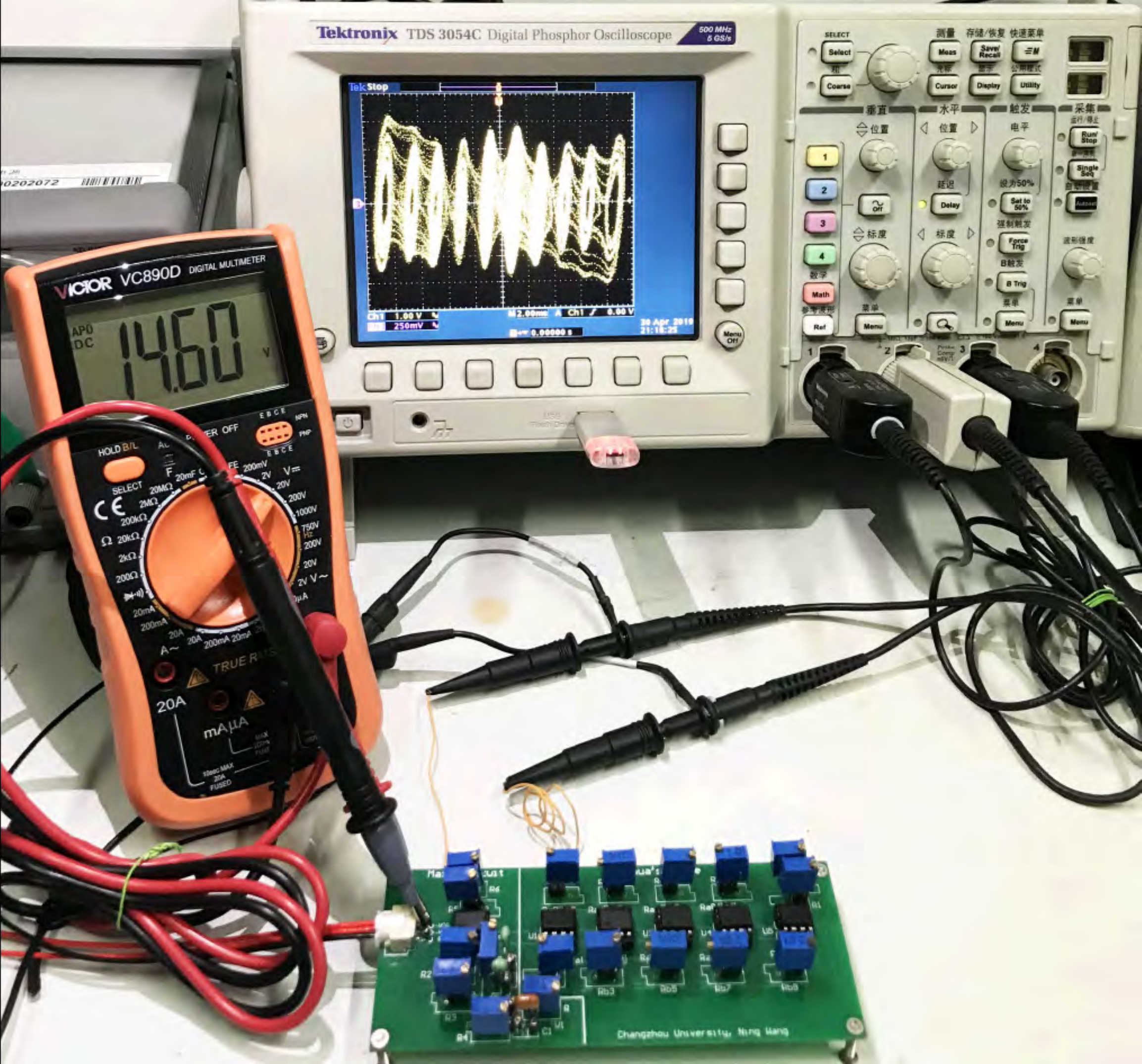}
\caption{Snapshot of the hardware experiment setup.}
\label{fig:experimentalphoto}
\end{figure}

Using the circuit simulation model shown in Fig.~\ref{fig:Multisimcircuit}, the Sallen-Key HPF-based Chua's circuit with the simplified multi-segment piecewise-linear Chua's diode was performed on a printed circuit board, where adjustable resistors, monolithic ceramic capacitors, and op-amps TL082CP with \SI{\pm14.6}{\volt} DC power supplies ($\pm E_{\rm sat}\approx \SI{\pm13}{\volt}$) were adopted. The snapshot of the hardware breadboard linked with a digital oscilloscope is shown in Fig.~\ref{fig:experimentalphoto}.

\begin{figure}[!htb]
	\centering	
	\begin{minipage}{\TwoImW}
		\centering
		\includegraphics[keepaspectratio,width=\TwoImW,height=\TwoImW]{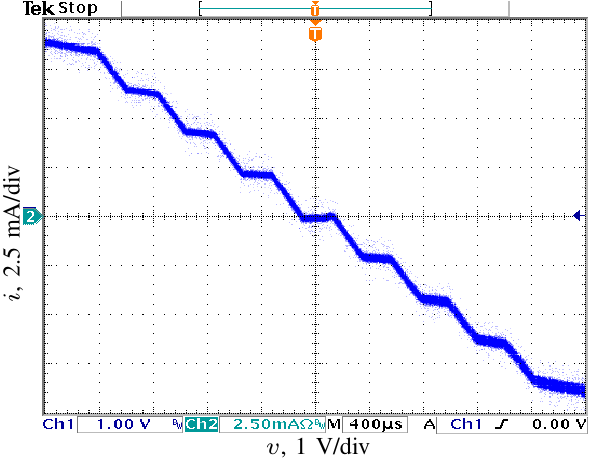}
		\subcaption*{(a)}
	\end{minipage}
	\begin{minipage}{\TwoImW}
		\centering
		\includegraphics[keepaspectratio,width=\TwoImW,height=\TwoImW]{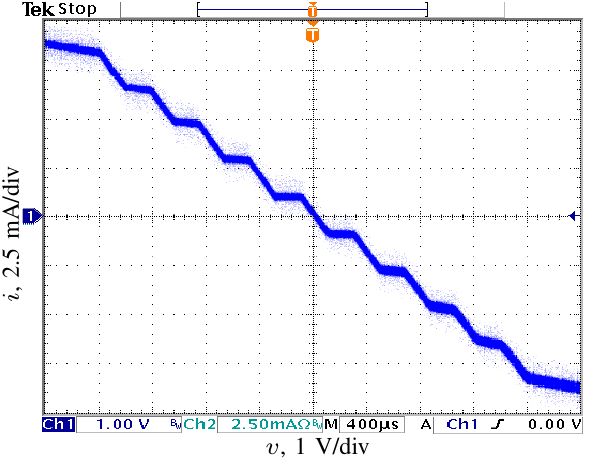}
		\subcaption*{(b)}
	\end{minipage}\\
	\begin{minipage}{\TwoImW}
		\centering
		\includegraphics[keepaspectratio,width=\TwoImW,height=\TwoImW]{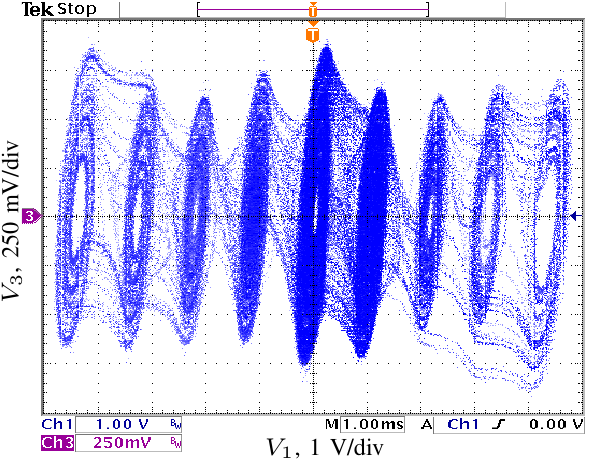}
		\subcaption*{(c)}
	\end{minipage}
	\begin{minipage}{\TwoImW}
		\centering
		\includegraphics[keepaspectratio,width=\TwoImW,height=\TwoImW]{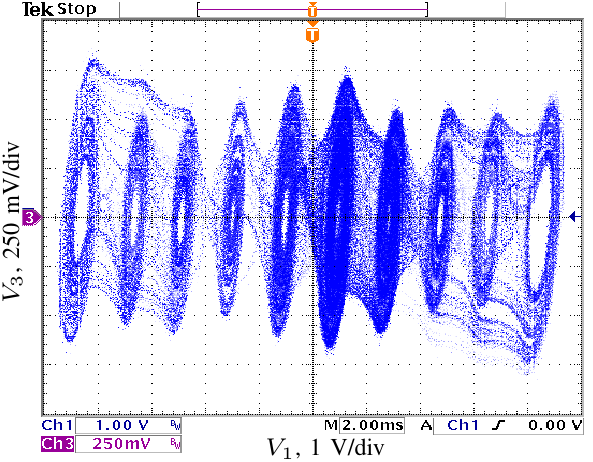}
		\subcaption*{(d)}
	\end{minipage}
	\caption{The results of hardware experiments: (a) the 17-segment piecewise-linear curve; (b) the 19-segment piecewise-linear curve; (c) the 9-scroll chaotic attractor; (d) the 10-scroll chaotic attractor.}
	\label{fig:experimentalresults}
\end{figure}

To verify the $v-i$ characteristics of the proposed Chua's diode, the continuous sinusoidal signal
\[
v_s=V_{\rm m}{\rm sin}(2\pi ft)
\]
provided by Tektronix AFG3022 function generator and Tektronix TCP213A current probe were used for the driven voltage and to detect the output current,
respectively. The experimental results were intuitively captured by Tektronix TDS3054C (a four-channel digital phosphor oscilloscope). When the continuous sinusoidal signal $v_s$ is deployed as $V_{\rm m}=5$V and $f=\SI{100}{\hertz}$, the measured loci in the $v-i$ plane are shown in Fig.~\ref{fig:experimentalresults}(a) and (b), respectively. Note that the captured currents are magnified four times for a better view. Furthermore, the phase portraits of 9-scroll and 10-scroll chaotic attractors in the $V_1-V_3$ plane are measured and presented in Fig.~\ref{fig:experimentalresults}(c) and (d), respectively.

Ignoring the deviations caused by parasitic circuit parameters, one can see that
the circuit simulations and hardware experiments are both consistent well with that obtained via numerical simulations, which demonstrates the feasibility of the proposed scheme.

However, the VOA has an inherent disadvantage, e.g. the relatively narrow frequency spectrum scope. To a certain extent, this shortcoming restricts the practical applications. In contrast, the current feedback op-amp, e.g. AD844, has wider bandwidth and exhibits higher response speed \cite{Zuo2014High,Lin2016Generation}. The AD844 can be further used to improve high frequency performance of Chua's circuit \cite{R1998Implementation}. Besides, the existing work \cite{Jin2018Low} has provided convincing evidence that the proposed inductor-free circuit modes can be designed by integrated circuits. The design and implementation of the low-voltage low-power multi-scroll integrated circuit deserve further attention with progress of the related engineering and electronic technology \cite{Trejo2012Integrated,Jin2018Low,Zhang2019A}.

\section{Conclusion}

In this paper, a simple scheme synthesizing a multi-segment piecewise-linear Chua's diode with simplified circuit implementation was introduced. Using the proposed Chua's diode, a Sallen-Key HPF-based Chua's circuit was further designed to generate multi-scroll chaotic attractors. The number of scrolls can be regulated by modifying the number of basic circuit cells and the corresponding parameters of the proposed Chua's diode.
Another important feature is that the dynamical transitions and different scroll types are adjustable with variation of two coupled parameters when the parameters of Sallen-Key HPF and Chua's diode are fixed.
Basic dynamical behaviors of the chaotic circuit were investigated via theoretical analysis and numerical simulation. The performance of the proposed scheme was verified by a large number of experiments performed on both multisim circuit design software and hardware platforms. An application prototype of image encryption was
established with the pseudorandom bit sequences generated from the proposed system. To sum up, the proposed design scheme has multiple advantages: 1) the multi-piecewise Chua's diode can be implemented by using modular circuit cells with systematically configured parameters; 2) the modified Chua's circuit has lower implementation complexity; 3) the proposed multi-scroll Chua's circuit is inductor-free and fully autonomous without external forcing; 4) the attractors can be generated from the circuit with different power supplies. Implementing the proposed scheme with integrated circuit deserves further investigation.

\section*{Acknowledgement}

The authors would like to thank the Associate Editor and the three anonymous reviewers for their valuable comments and suggestions that greatly contribute to improving the quality of the manuscript.

\bibliographystyle{IEEEtran_doi} 
\bibliography{Diode}


\graphicspath{{author_figures_pdf/}}

\begin{IEEEbiography}[{\includegraphics[width=1in,height=1.25in,clip,keepaspectratio]{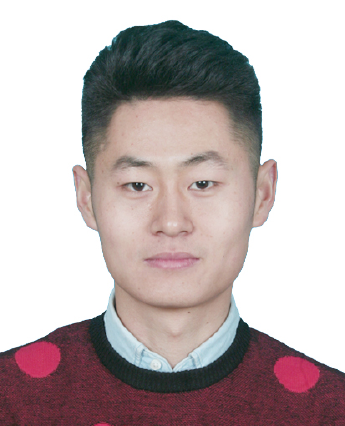}}]{Ning Wang}
    received the B.Sc. and M.Sc. degree in electronic information engineering and computer application technology from Changzhou University, China, in 2015 and 2018, respectively.
    He is currently pursuing the Ph.D. degree in control science and engineering at Tianjin University, China.

    His research interests include nonlinear circuits and systems, chaotic circuits and systems.
\end{IEEEbiography}

\vspace{-10mm}	

\begin{IEEEbiography}[{\includegraphics[width=1.1in, height=1.25in,clip,keepaspectratio]{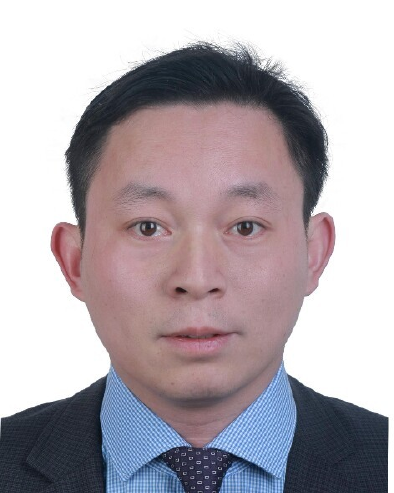}}]{Chengqing Li}(M'07-SM'13)
	received his M.Sc. degree in applied mathematics from Zhejiang University, China in 2005 and
	his Ph.D. degree in electronic engineering from City University of Hong Kong in 2008. Thereafter,
	he worked as a Postdoctoral Fellow at The Hong Kong Polytechnic University till September 2010.
	Then, he worked at the College of Information Engineering, Xiangtan University, China. From April 2013 to July 2014, he worked at the
	University of Konstanz, Germany, under the support of the Alexander von Humboldt Foundation.
	Since April 2018, he has been working with the School of Computer Science and Electronic Engineering, Hunan University, China as a full professor.
	
	Prof. Li focuses on dynamics analysis of digital chaotic systems and their applications in multimedia security.
	He has published more than fifty papers on the focal subject in the past 15 years, receiving more than 2800
	citations with h-index 29.
\end{IEEEbiography}

\vspace{-10mm}	

\begin{IEEEbiography}[{\includegraphics[width=1in,height=1.25in,clip,keepaspectratio]{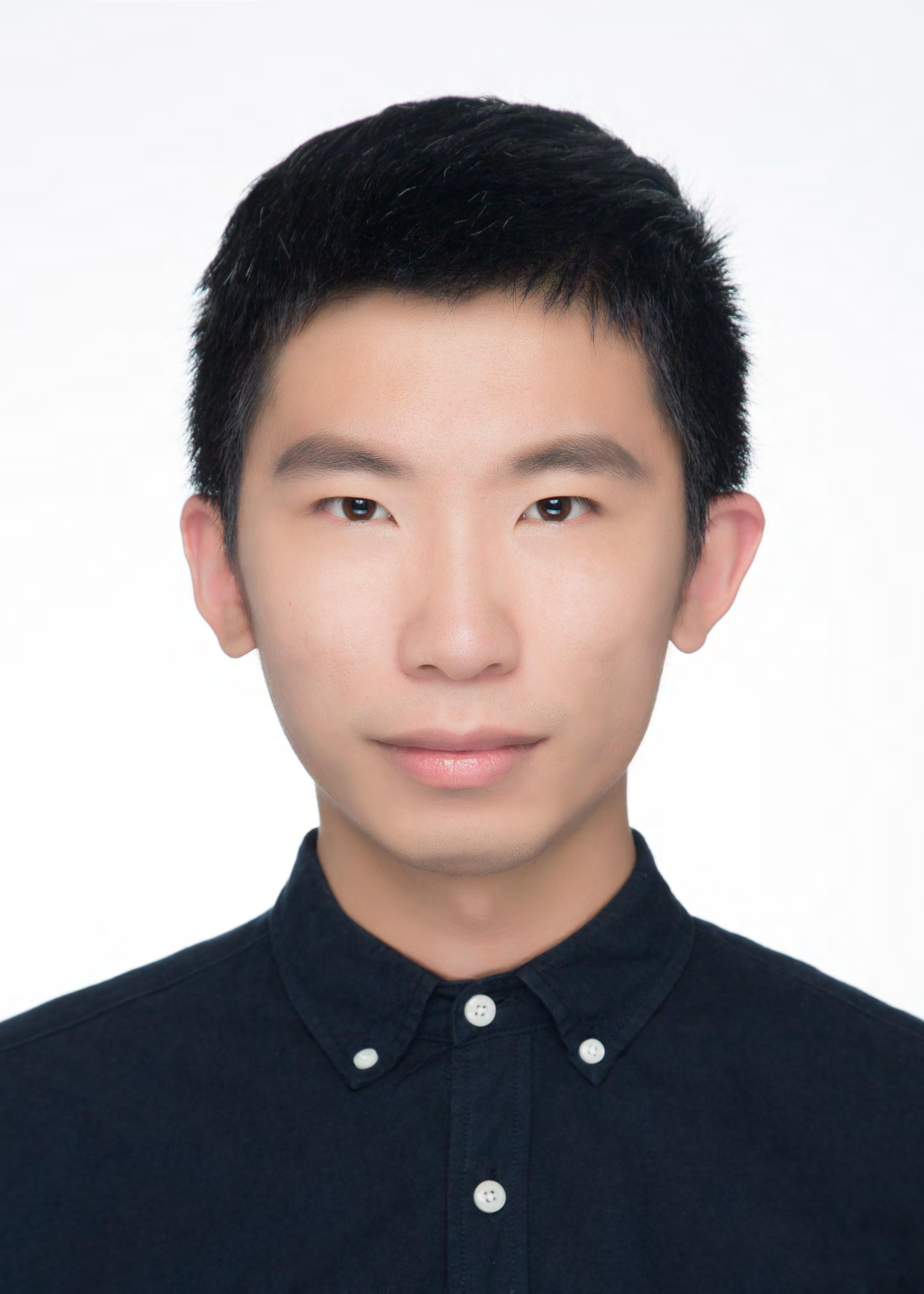}}]{Han Bao}
    received the B.Sc. degree from Finance and Economics University of Jiangxi, China, in 2015 and the M.Sc. degree from Changzhou University, China, in 2018.
    He is currently pursuing the Ph.D. degree in nonlinear system analysis and measurement technology at Nanjing University of Aeronautics and Astronautics, China.

    His research interests include memristive neuromorphic circuit, computer science, and artificial intelligence.
\end{IEEEbiography}

\begin{IEEEbiography}[{\includegraphics[width=1in,height=1.25in,clip,keepaspectratio]{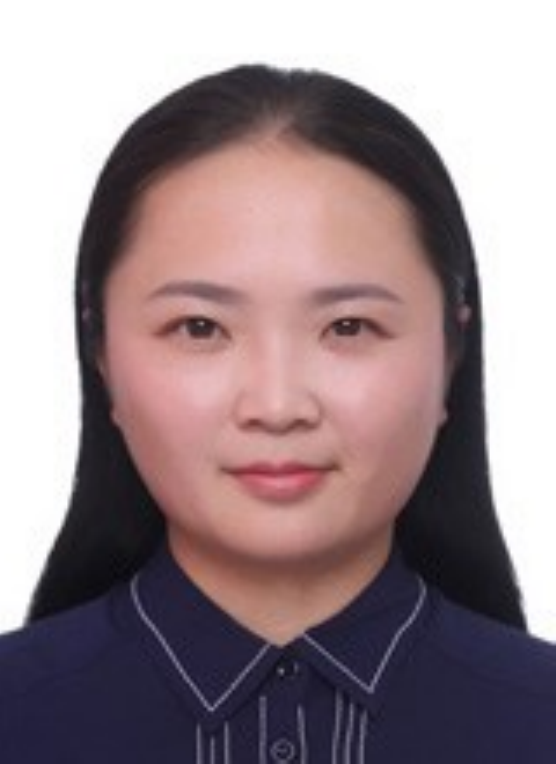}}]{Mo Chen}
    received the B.S. degree in information engineering, and the M.S. and Ph.D. degrees in Electromagnetic Field and Microwave Technology from Southeast University, Nanjing, China, in 2003, 2006 and 2009, respectively. From March 2009 to July 2013, she was a Lecturer in Southeast University, Nanjing, China. She is currently an associate professor in the School of Information Science and Engineering, Changzhou University, Changzhou, China.

    Her research interest mainly focuses on memristor and its application circuits, and nonlinear circuits and systems.
\end{IEEEbiography}


\begin{IEEEbiography}[{\includegraphics[width=1in,height=1.25in,clip,keepaspectratio]{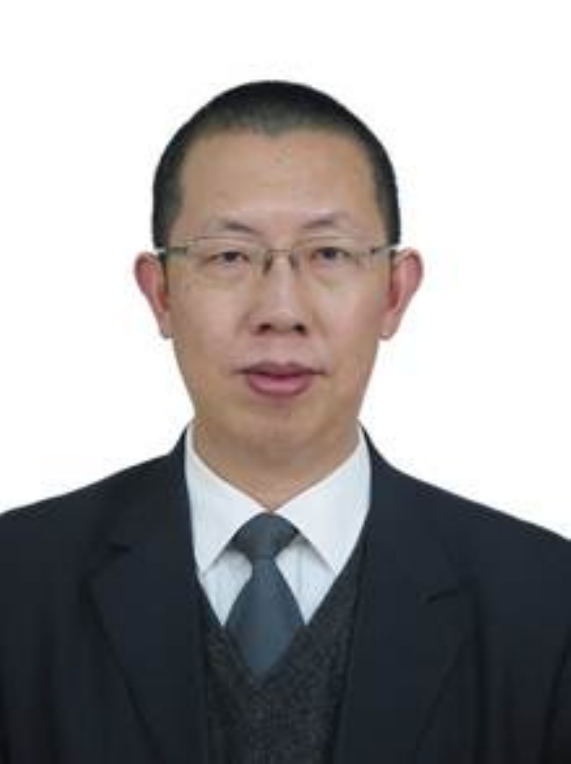}}]{Bocheng Bao}
    received the B.S. and M.S. degrees in electronic engineering from the University of Electronics Science and Technology of China, Chengdu, China, in 1986 and 1989, respectively, and the Ph.D. degree with the Department of Electronic Engineering, Nanjing University of Science and Technology, Nanjing, China, in 2010.

    He has more than 20 years' experience in industry and has ever been on several enterprises serving as Senior Engineer and General Manager.
    From June 2008 to January 2011, he was a Professor in the School of Electrical and Information Engineering, Jiangsu University of Technology, Changzhou, China.
    He is currently a Professor in the School of Information Science and Engineering, Changzhou University, Changzhou, China.
    His research interests include bifurcation and chaos, analysis and simulation in neuromorphic circuits, power electronic circuits, and nonlinear circuits and systems.
\end{IEEEbiography}
\vfill

\end{document}